\documentclass[journal]{IEEEtran}
\usepackage{geometry}
\usepackage{ifpdf}

\usepackage{cite}
\usepackage{epstopdf}
\usepackage{booktabs}
\usepackage{multirow}
\usepackage{graphicx}
\usepackage{longtable}
\usepackage{verbatim}
\usepackage{color}
\usepackage{xcolor}
\usepackage{subcaption}
\usepackage[normalem]{ulem} 
\newcommand\hl{\bgroup\markoverwith
  {\textcolor{yellow}{\rule[-0.5ex]{2pt}{2.5ex}}}\ULon}

\usepackage[cmex10]{amsmath}
\usepackage{amsmath,bm}
\usepackage{amsfonts,amssymb}

\usepackage{algorithmic}

\usepackage{array}

\usepackage{setspace}

\usepackage{url}
\usepackage[hidelinks]{hyperref}

\begin{document}

%
\title{A Lightweight, Transferable, and Self-Adaptive Framework for Intelligent DC Arc-Fault Detection in Photovoltaic Systems}

%
%
%

\author{Xiaoke~Yang, 
        Long~Gao, 
        Haoyu~He,
        Hanyuan Hang, 
        Qi Liu,
        Shuai~Zhao,
        Qiantu~Tuo, 
        and Rui~Li

\thanks{This work was supported by EcoFlow Research Projects. (\emph{Corresponding author: Rui Li})}.
\thanks{ X. Yang, L. Gao, H. Hang, Q. Liu, Q. Tuo, and R. Li are with EcoFlow Inc, Shenzhen, China. (Email: \{xiaoke.yang, leo.gao, haoyu.he, hans.hang, thomas.edison, ricky.li\}@ecoflow.com).}
\thanks{S. Zhao is with AAU Energy, Aalborg University, Denmark (Email: szh@energy.aau.dk).}
}

\maketitle

\begin{abstract}
Arc-fault circuit interrupters (AFCIs) are essential for mitigating fire hazards in residential photovoltaic (PV) systems, yet achieving reliable DC arc-fault detection under real-world deployment conditions remains challenging.
Spectral interference from inverter switching behavior, hardware heterogeneity across power converters, long-term operating-condition drift, and environmental noise collectively compromise the robustness of conventional AFCI solutions.
This paper proposes a lightweight, transferable, and self-adaptive learning-driven (LD) framework  for intelligent DC arc-fault detection in PV systems.
At the device level, a spectrum-based neural network, \textit{LD-Spec}, learns compact spectral representations that enable efficient on-device inference and near-perfect arc discrimination.
Across heterogeneous inverter platforms, \textit{LD-Align} performs cross-hardware representation alignment to ensure robust detection performance despite hardware-induced distribution shifts.
To address long-term field evolution, \textit{LD-Adapt} introduces a cloud--edge collaborative self-adaptive updating mechanism that detects previously unseen operating conditions and performs controlled model evolution.
Extensive hardware experiments 
demonstrate near-perfect detection performance, achieving an accuracy of 0.9999 and an F1-score of 0.9996.
Across diverse nuisance-trip-prone operating conditions, including inverter start-up, grid transitions, load switching, and harmonic grid disturbances, the proposed method achieves a 0\% false-trip rate.
Cross-hardware transfer experiments further show that reliable adaptation can be achieved using only 0.5\%--1\% labeled target data while preserving source-domain performance.
Field adaptation experiments demonstrate that the proposed adaptive mechanism can recover detection precision from 21\% to 95\% under previously unseen operating conditions.
These results indicate that the proposed LD-framework enables a scalable and deployment-oriented AFCI solution capable of maintaining highly reliable arc-fault detection across heterogeneous devices and long-term real-world operation.
\end{abstract}


\IEEEpeerreviewmaketitle

\section{Introduction}

\IEEEPARstart{R}{esidential} and commercial photovoltaic (PV) systems integrated with battery energy storage systems (BESS) are being rapidly deployed as module costs decline and distributed generation expands in low-voltage networks~\cite{Saxena2018_TIE,IEA_WEO2024}. In such power-electronic infrastructures, DC-side \emph{series arc faults} remain a critical safety hazard, causing fire risk, equipment damage, and service interruption. Unlike AC parallel arcs, series DC arcs are sustained by continuous current flow and exhibit stochastic plasma behavior and broadband spectral signatures, making reliable detection difficult under practical operating conditions~\cite{Georgijevic2016_TPEL, Ahn2023_TIE}. These risks have motivated stringent safety requirements such as UL~1699B and widespread deployment of arc-fault circuit interrupters (AFCIs).

Early AFCI methods mainly relied on threshold rules or handcrafted physics-inspired indicators derived from current magnitude, spectral energy, or time--frequency statistics. Although computationally efficient, such methods degrade under inverter switching harmonics, maximum power point tracking (MPPT) perturbations, and dynamic load variations. As converter switching frequency and control complexity increase, the spectral background becomes highly nonstationary, further weakening fixed decision criteria~\cite{Xiong2018_SolarEnergy,Ahn2024_TII}. More recently, lightweight convolutional models and time--frequency representations have been introduced to improve arc discrimination~\cite{Paul2025_TPEL,Yin2025_TPEL}. However, most existing results are obtained under controlled laboratory conditions and assume a fixed inverter platform.

Practical AFCI deployment faces three fundamental challenges. \emph{First, intra-system variability}: MPPT perturbations, inverter mode transitions, grid impedance fluctuations, and load dynamics introduce broadband disturbances that overlap with arc signatures, often causing nuisance trips or missed detections in real PV systems~\cite{Omran2022_IEEES,Kim2024_TPEL}. \emph{Second, cross-converter heterogeneity}: differences in switching frequency, semiconductor technology, modulation strategy, and filter topology systematically reshape measured spectra; EMI studies confirm that hardware-dependent switching dynamics strongly affects frequency-domain characteristics~\cite{Trzynadlowski2011_PowerElectronicsHandbook}. Since arc-fault data collection is hazardous and costly, practical models must generalize across heterogeneous converters with limited target supervision, motivating domain adaptation and invariant representation learning~\cite{Ganin2016_ICLR,pmlr-v235-lai24c,pmlr-v119-ahuja20a,pmlr-v97-zhang19i}. \emph{Third, long-term temporal drift}: seasonal irradiance variation, temperature-dependent behavior, connector aging, and sensor degradation gradually shift baseline spectra; long-term monitoring studies have documented such drift in deployed PV systems~\cite{Jordan2015_RSER,Lindig2021_ProgressPhotovoltaics}. Without adaptive updating, static AFCI models may suffer degraded detection performance or elevated false-alarm rates. Continual learning studies further suggest that controlled updating is essential under nonstationary environments~\cite{Kirkpatrick2017_PNAS}.

These observations indicate that reliable AFCI performance is not an intrinsic property of a classifier trained on one converter, but depends on maintaining spectral separability, stable error behavior, and operating-condition consistency across devices and over time. At the same time, practical AFCI implementation must satisfy strict industrial constraints, including millisecond-level latency and microcontroller-scale memory budgets. Therefore, a deployment-oriented AFCI solution must jointly provide robust on-device detection, cross-hardware transferability, and long-term field adaptability.

To address these challenges, we propose a lightweight, transferable, and self-adaptive \emph{LD-framework} for lifecycle-aware AFCI protection in PV--BESS systems. The framework integrates three components. \emph{LD-Spec} is a microcontroller-efficient spectral backbone that learns compact arc-discriminative frequency structure for robust on-device detection. \emph{LD-Align} performs cross-converter representation alignment to gain robust detection under hardware-induced distribution shifts. \emph{LD-Adapt} enables cloud--device collaborative self-evolution through operating-condition novelty detection and lightweight safety-aware model updating. Together, these components form a closed-loop deployment pipeline in which field devices perform local detection and monitoring, cloud-side adaptation extracts transferable knowledge across the fleet, and updated models are delivered through secure OTA updates.

The main contributions of this work are summarized as follows:

\begin{itemize}


\item[\textit{(i)}] \textit{Transfer-aware spectral AFCI framework:} We propose LD-Align, a cross-hardware transfer framework built upon the existing spectrum-based backbone, LD-Spec. Leveraging LD-Spec's ability to extract arc-discriminative spectral features under deployment-oriented constraints, LD-Align preserves arc-relevant spectral structures and mitigates hardware-induced distribution shifts with minimal target supervision, all while satisfying embedded latency and memory requirements.

\item[\textit{(ii)}] \textit{Lifecycle-aware cloud-device adaptive learning:}
We propose LD-Adapt, a cloud--device coordination mechanism for novel data samples detection and controlled model evolution, enabling stable and safety-aware AFCI updates under long-term field drift.

\item[\textit{(iii)}] \textit{Industrial-scale validation:}
We conduct comprehensive experiments across multiple inverter frequencies, nuisance-trip-prone operating categories, UL~1699B secnarios, cross-platform transfers, scaling-law analysis, and real-world operating-condition shifts, demonstrating sustained detection reliability under hardware heterogeneity and lifecycle evolution.
\end{itemize}

The remainder of this paper is organized as follows. Section~\ref{sec:problem} formulates the system model and problem statement. Section~\ref{sec:method} details the proposed LD-framework. Section~\ref{sec:experiments} presents the experimental results. Section~\ref{sec:conclusion} concludes this paper.

\section{System Description and Challenges}
\label{sec:problem}

This section describes residential photovoltaic (PV) and battery energy storage systems (BESS) from both system-level and signal-level perspectives, and analyzes the physical, spectral, and embedded-computing constraints that fundamentally affect DC arc-fault detection. 
These factors lead to three structural challenges that motivate the proposed learning-driven framework.

\subsection{System Architecture}
\begin{figure}[htbp]
\centering
\includegraphics[width=0.95\linewidth]{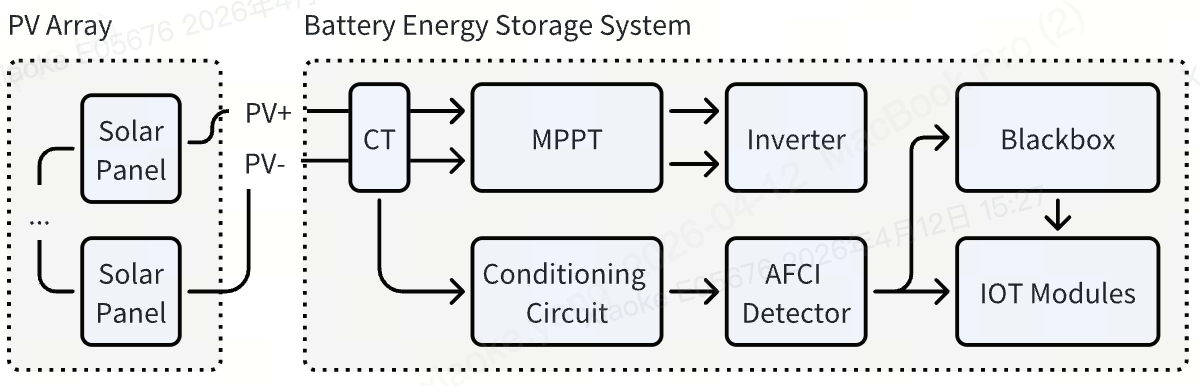}
\caption{Overview of the PV-BESS system architecture.}
\label{fig:system_overview}
\end{figure}

The studied system is illustrated in Fig.~\ref{fig:system_overview}. It follows a typical residential PV--BESS configuration, in which PV modules deliver power to a inverter via an MPPT stage, enabling regulated power exchange with the grid.
Series DC arc faults typically originate along PV strings, particularly at connectors and junctions, components that are susceptible to mechanical wear, humidity ingress, corrosion, and thermal cycling.

As shown in Fig.~\ref{fig:system_overview}, arc-related current signals are measured at the PV input using current transformers (CTs), conditioned 
 by analog front-end circuitry, then digitized and processed on-device by an embedded processor for real-time AFCI inference. Although the processor integrates a neural processing unit (600 MOPS on INT8), limited memory (133 kB) and flash capacity (1MB), coupled wclaude config add allowedTools "Shell(*)"
ith restricted kernel support, impose strict constraints on model size, memory usage, and computational complexity. These constraints significantly restrict feasible model designs and motivate compact frequency-domain representations and lightweight inference architectures.

\subsection{DC Arc-Fault Physics and Signal Characteristics}
\begin{figure}[!htbp]
\centering
\includegraphics[width=0.95\linewidth]{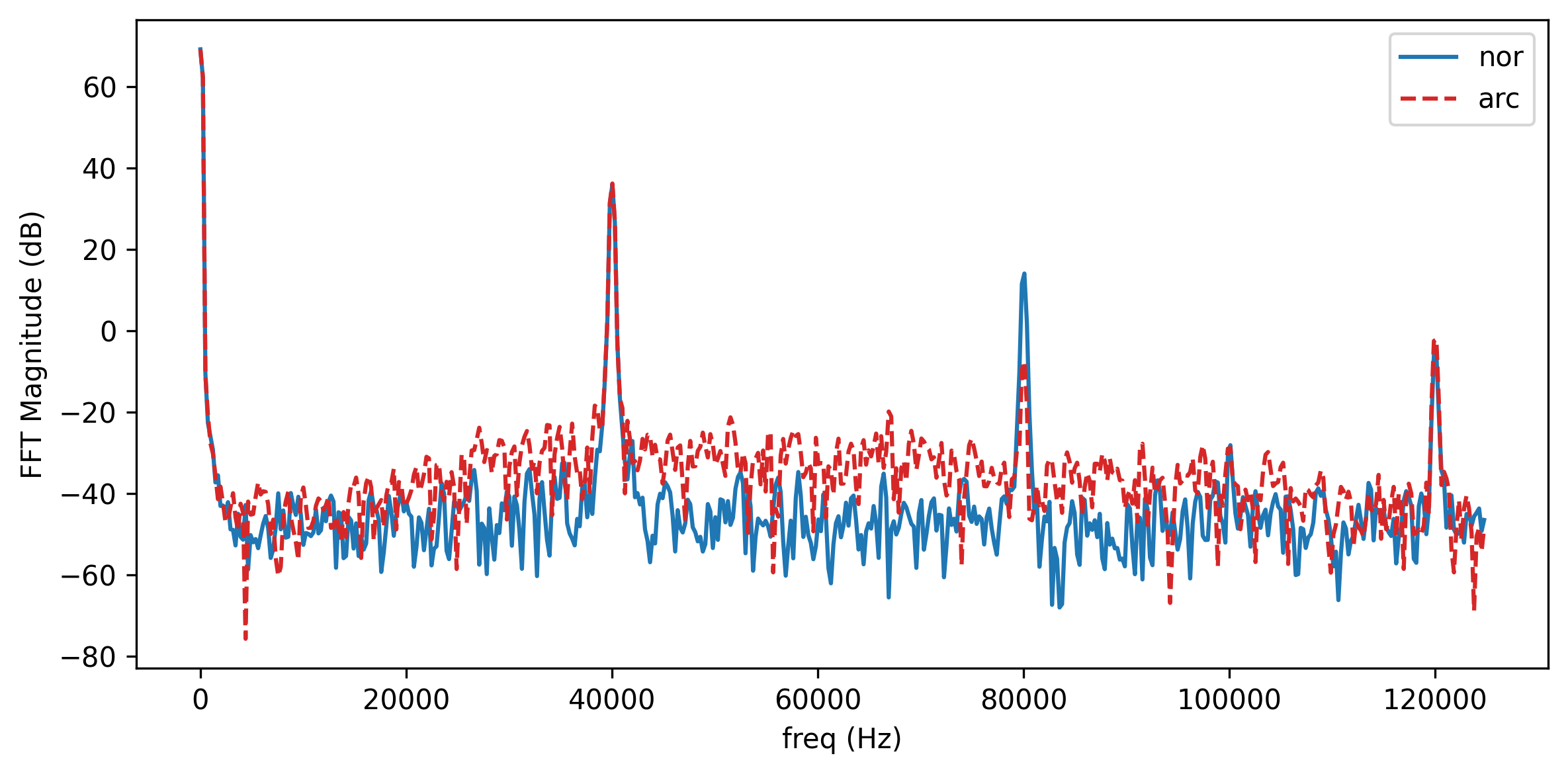}
\caption{Spectrum of PV currents during normal and arc-fault operation.
Arc faults exhibit elevated broadband spectral magnitude.}
\label{fig:normal_vs_arc}
\end{figure}

DC arc faults originate from ionized air gaps caused by connector degradation, insulation aging, corrosion, or mechanical fatigue. From a signal perspective, DC arc faults exhibit three distinct frequency-domain characteristics: (i) broadband high-frequency energy elevation, (ii) stochastic and impulsive temporal bursts, and (iii) spectral energy redistribution, as illustrated in Fig.~\ref{fig:normal_vs_arc}.

Under practical operating conditions, these arc-induced components often overlap with inverter switching and control-related spectral artifacts, particularly under varying irradiance, load, and operating modes.
Such spectral overlap significantly limits the effectiveness of traditional threshold-based detectors and motivates learning-based spectral discrimination under strict embedded constraints.

\subsection{Challenges of Reliable Arc-Fault Detection}
\label{sec:challenges}

\begin{figure}[!htbp]
\centering
\begin{subfigure}{0.95\linewidth}
    \includegraphics[width=\linewidth]{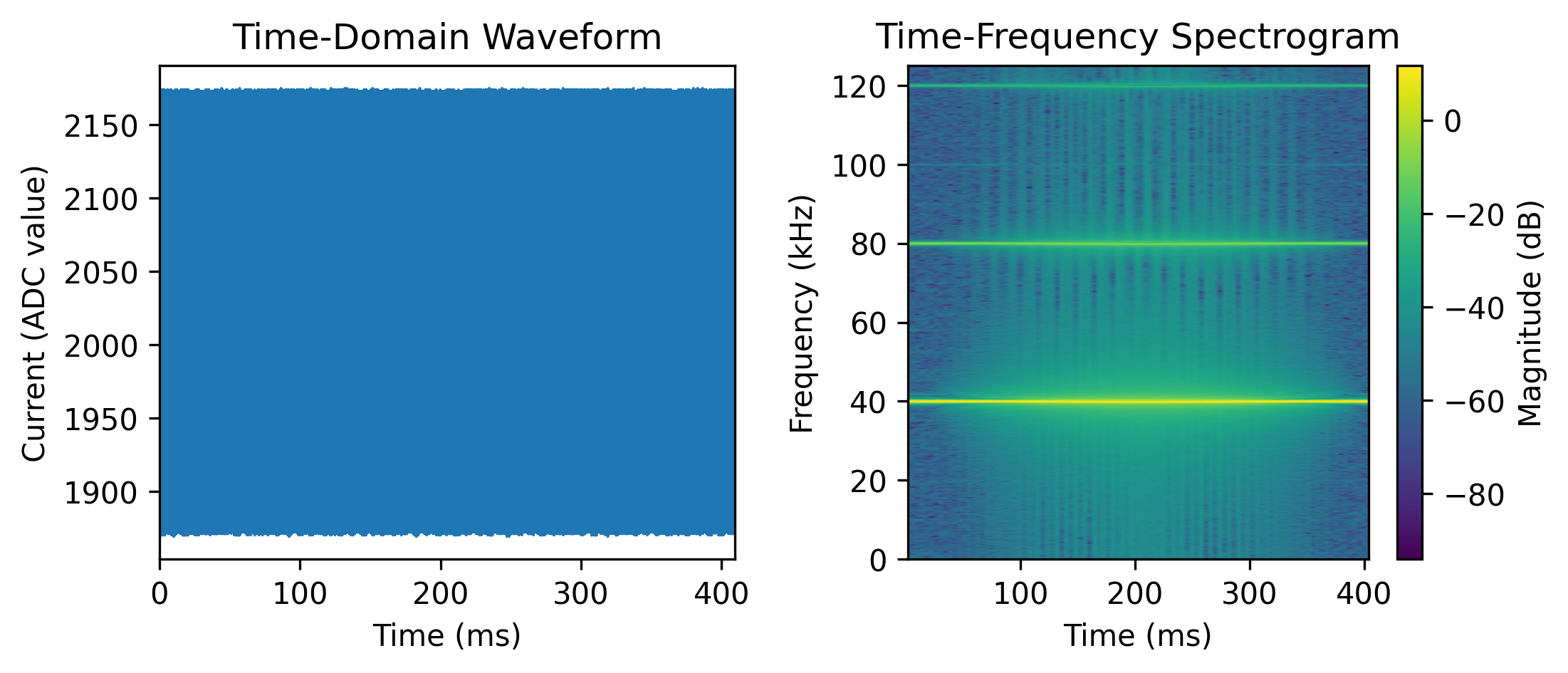}
    \caption{(a) PV current under steady operation.}
\end{subfigure}
\vspace{0.35cm}
\begin{subfigure}{0.95\linewidth}
    \includegraphics[width=\linewidth]{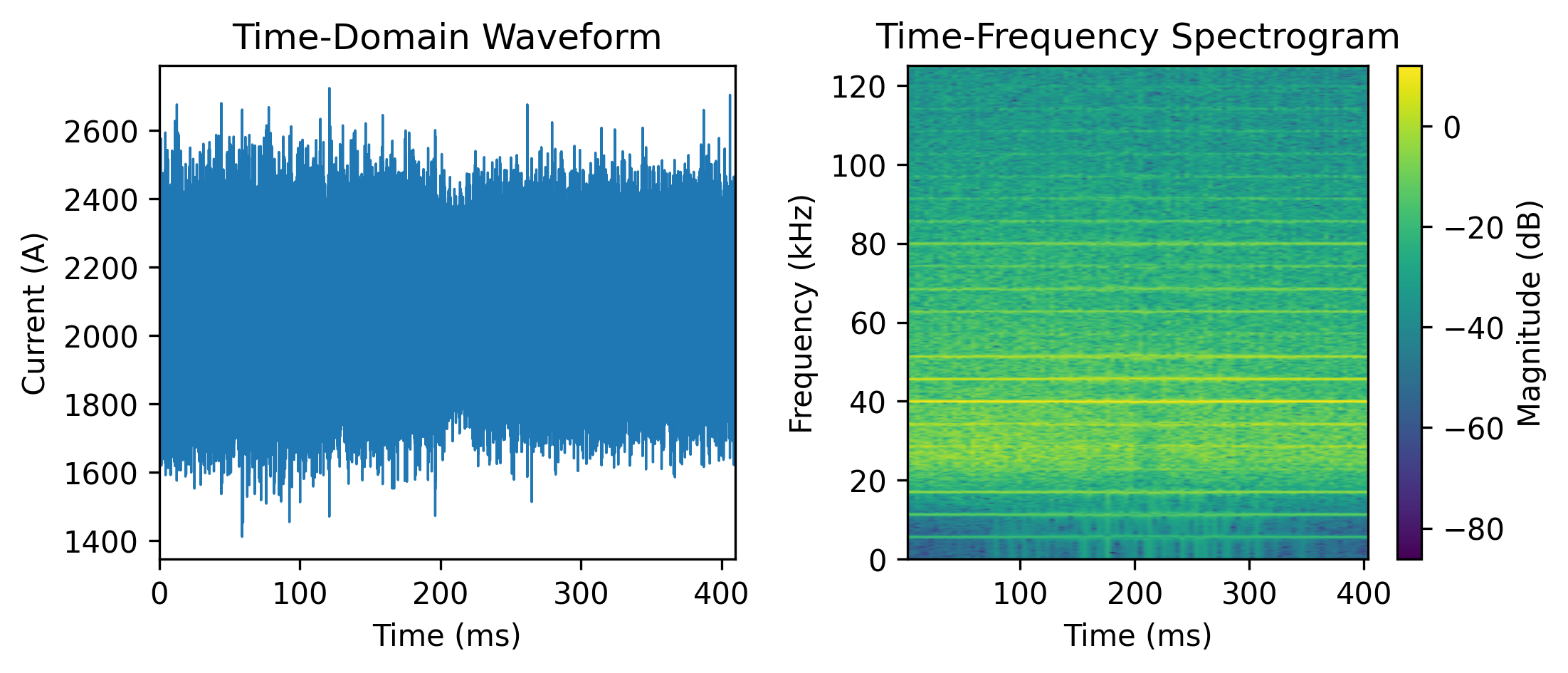}
    \caption{(b) PV current under arc fault.}
\end{subfigure}
\caption{Time-domain waveforms and spectrograms of PV current under steady operation and arc-fault conditions.}
\label{fig:arc_fault_examples}
\end{figure}

\paragraph{Challenge~1: \textit{Intra-System Variability}}
Residential PV-BESS systems operate under highly dynamic conditions. MPPT perturbations, grid synchronization events, relay switching, AC/DC load steps, and inverter mode transitions introduce broadband spectral disturbances even during normal operation. While steady-state operation (Fig.~\ref{fig:arc_fault_examples}(a) and (b)) exhibits structured harmonic patterns, arc faults (Fig.~\ref{fig:arc_fault_examples}(b)) produce broadband spectral energy across a wide frequency range. This 
intra-system spectral drift necessitates \emph{compact yet highly discriminative} representations that remain robust across operating conditions, motivating the proposed LD-Spec module.

\begin{figure}[htbp]
\centering
\includegraphics[width=0.45\textwidth]{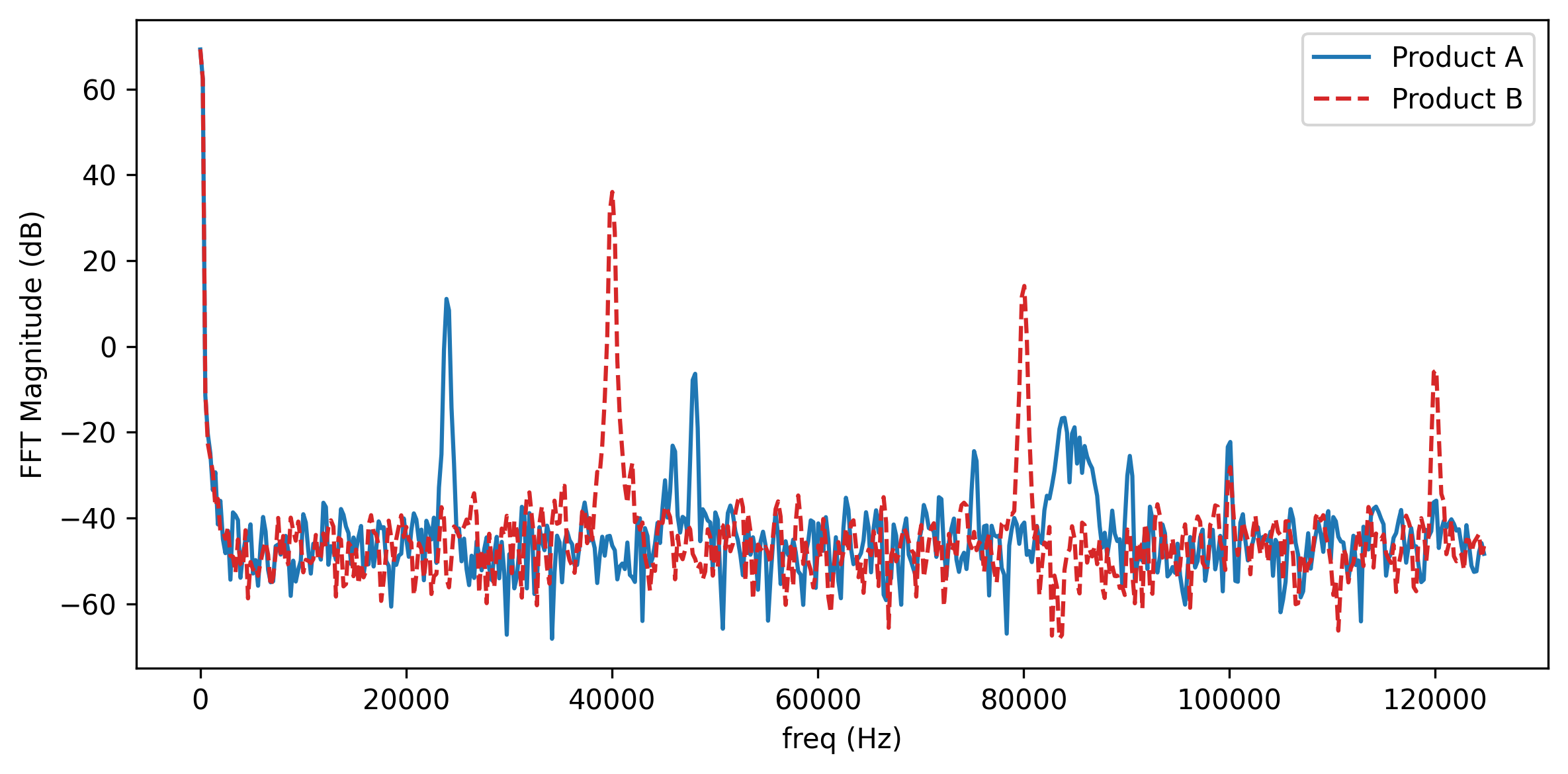}
\caption{Spectrum of PV currents from two inverter platforms exhibiting distinct switching-frequency characteristics.}
\label{fig:series_frequency_diff}
\end{figure}

\paragraph{Challenge~2: \textit{Cross-Hardware Heterogeneity}}
Inverter hardware differs significantly in switching frequency, semiconductor technology, circuit topology, and control strategies.
These hardware differences produce platform-specific spectral distributions during normal operation, as illustrated in
Fig.~\ref{fig:series_frequency_diff}. Since collecting arc-fault data on every hardware platform is hazardous and costly, device-specific supervised training is impractical. Reliable AFCI deployment therefore requires cross-converter representation alignment to preserve arc discriminability under hardware transfer, motivating
the proposed LD-Align module.

\begin{figure}[!htbp]
\centering
\includegraphics[width=0.95\linewidth]{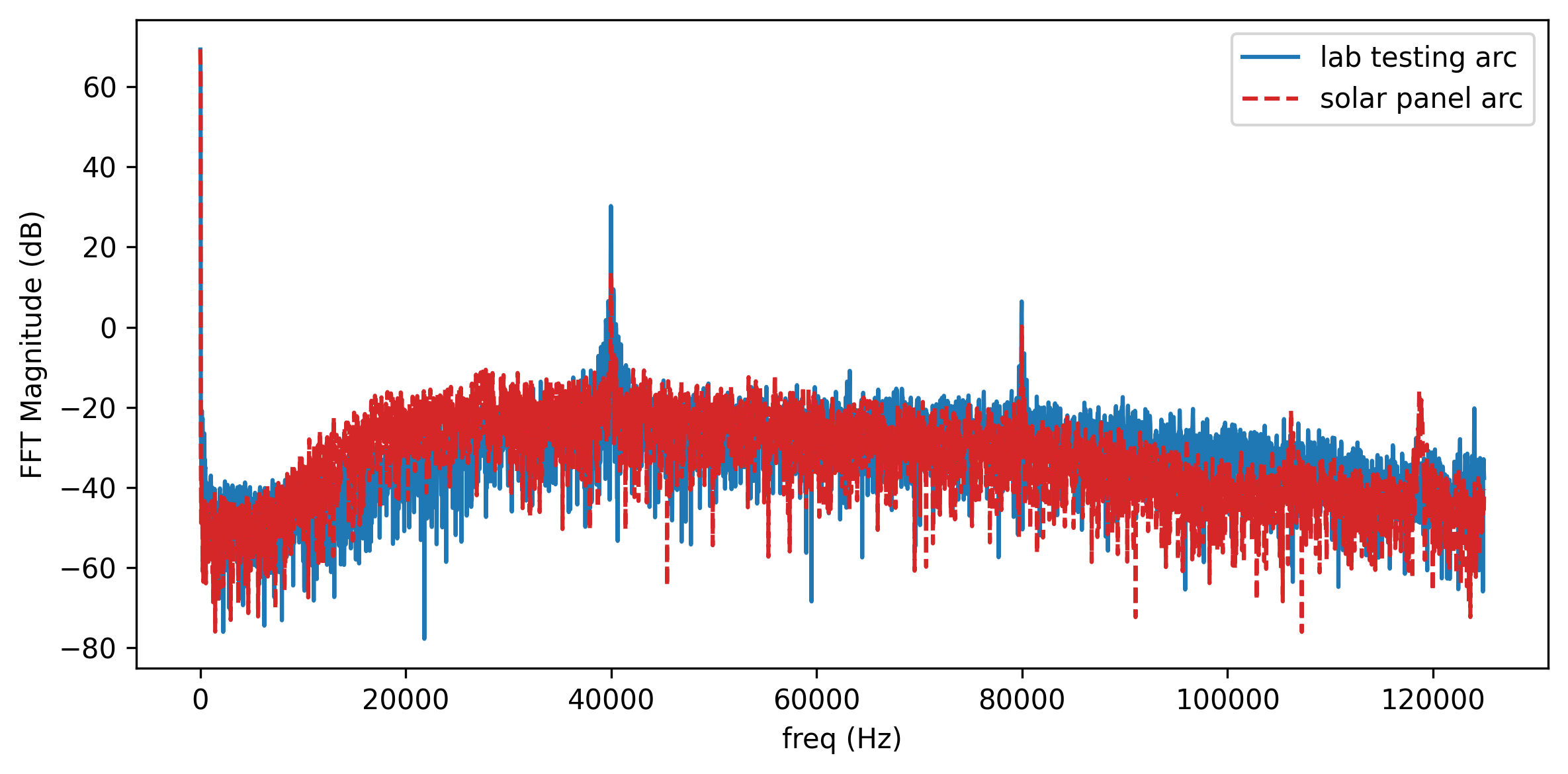}
\caption{Spectrum of PV currents from a laboratory PV source and a real solar
panel under identical current conditions.}
\label{fig:pv_source_vs_panel_spectro}
\end{figure}

\paragraph{Challenge~3: \textit{Real-World Deployment Drift}}
Even after extensive laboratory validation, field deployments often operate under conditions that differ from controlled test environments.
A typical example is shown in Fig.~\ref{fig:pv_source_vs_panel_spectro}, where current spectra measured from a laboratory PV source and an actual solar panel differ under identical current levels. Over longer time scales, component aging, environmental variations, and installation-specific factors further shift signal statistics. Maintaining reliable AFCI performance therefore requires detecting emerging operating conditions and adaptively updating model parameters, which motivates the proposed cloud-device self-evolving mechanism, LD-Adapt.

In summary, reliable AFCI operation in PV-BESS systems is governed by three key factors: \emph{intra-system variability}, motivating LD-Spec for robust on-device detection; \emph{cross-system heterogeneity}, motivating LD-Align for cross-hardware transfer; and \emph{long-term deployment drift}, motivating LD-Adapt for lifecycle-aware adaptation.
These challenges collectively motivate the unified LD-framework presented in Section~\ref{sec:method}.

\section{The LD-Framework}
\label{sec:method}

\begin{figure}[!htbp]
\centering
\includegraphics[width=0.50\textwidth]{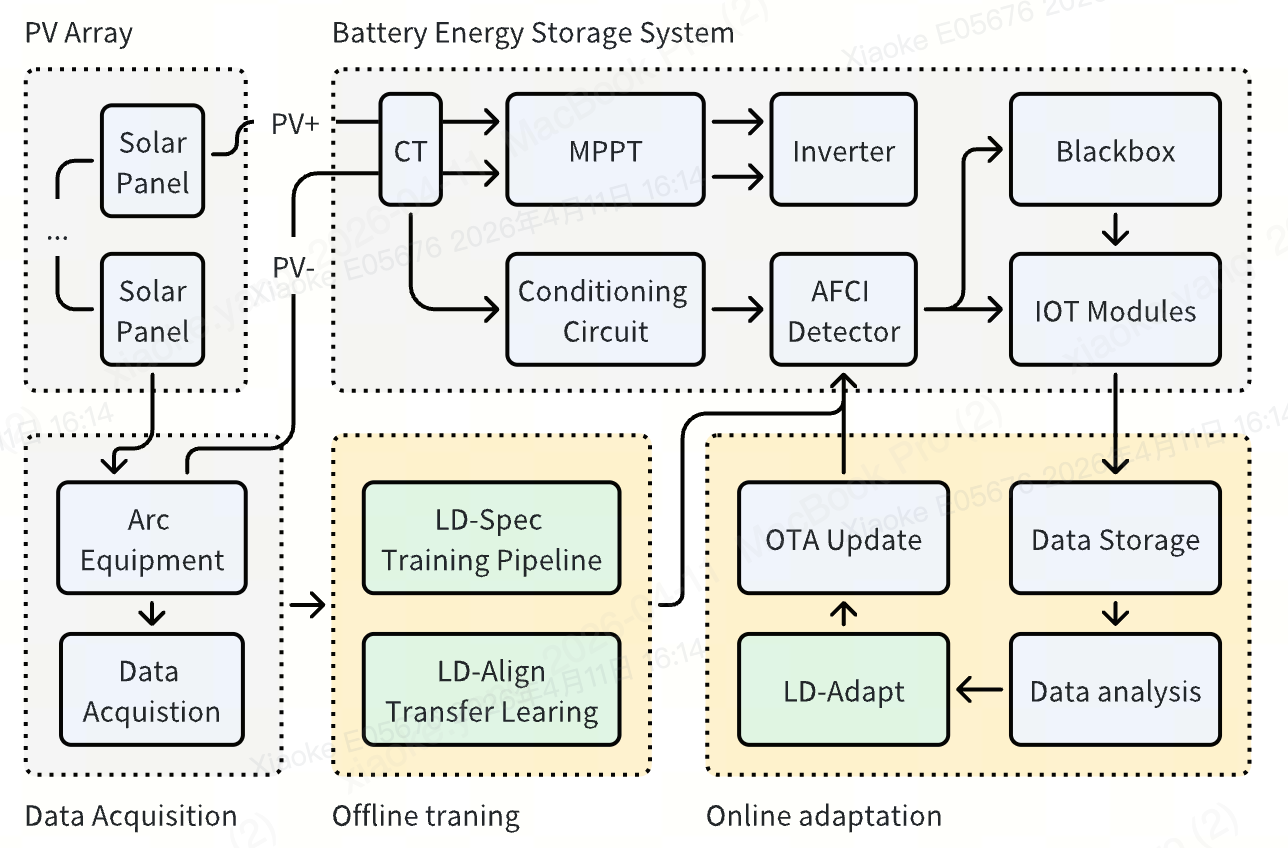}
\caption{System-level architecture of the proposed LD-framework, illustrating the interaction between on-device lightweight AFCI detection, offline data acquisition and training, cross-converter transfer learning (LD-Align), and online self-adaptive evolution (LD-Adapt) with OTA deployment.}
\label{fig:method_overview}
\end{figure}

Motivated by the challenges identified in Section~\ref{sec:challenges}, this section presents the \emph{LD-framework}, a learning-driven architecture for reliable DC arc-fault detection in residential PV-BESS systems, as shown in Figure~\ref{fig:method_overview}.
At the device level, each PV-BESS inverter hosts a lightweight AFCI
detection stack. The on-device module \emph{LD-Spec} performs real-time frequency-domain arc detection using high-speed current measurements acquired through current transformers (CTs) and conditioning circuits.
LD-Spec is the result of an offline data acquisition and training pipeline. Controlled arc-generation platforms and automated testing equipment are used to construct high-quality labeled datasets under laboratory conditions, establishing strong baseline separability between arc and non-arc spectral patterns before field deployment. Within this environment, \emph{LD-Align} performs cross-converter representation alignment to preserve arc-relevant spectral structure under hardware-induced distribution shifts.
Meanwhile, a cloud knowledge center aggregates spectral features, operating condition indicators, and evolution traces from deployed devices. \emph{LD-Adapt} then detects emerging operating conditions and updates models through controlled cloud-device evolution. Updated models or calibration parameters are subsequently delivered to
field devices via OTA deployment.

Through this coordinated cloud-device operation, the LD-framework forms
an adaptive feedback learning loop in which individual devices
evolve locally while benefiting from fleet-level knowledge sharing.
This design transforms arc-fault detection from a static laboratory
classifier into a \emph{maintainable, transferable, and deployment-ready}
capability capable of sustaining reliable AFCI protection under
hardware diversity and long-term operational drift.

\subsection{LD-Spec: Frequency-Domain Arc Detection}
\label{subsec:LD-Spec}
Reliable AFCI detection in PV-BESS systems must remain discriminative under diverse operating conditions, including MPPT start-up, steady-state operation, inverter mode transitions, load fluctuations, and other operating conditions, while satisfying strict embedded constraints. To address this requirement, we follow the end-to-end frequency-domain approach proposed in \cite{TI_TIDA-010955_2025} that integrates data acquisition, preprocessing, feature construction, and supervised model training.

\subsubsection{Feature Engineering}
DC arcs produce broadband, stochastic, and highly non-stationary current signatures that are difficult to isolate in the time domain. In contrast, frequency-domain representations naturally expose these patterns as energy elevation across mid- and high-frequency bands.
Accordingly, we adopt a frequency-centric feature engineering pipeline.

Current measurements are segmented into fixed-length frames
of $L$ samples:
\begin{align}
\boldsymbol{F}_i = \{x_t\}_{t=iL}^{(i+1)L-1}.
\label{eq:frame_segmentation}
\end{align}
A Hanning window $w[n] = 0.5 \left(1 - \cos\!\left({2\pi n}/(L-1)\right)\right)$ is then applied to reduce spectral leakage. Discrete Fourier Transform (DFT) is then performed on then windowed frame,
\begin{align}
X[k] = \sum_{n=0}^{L-1} x[n] e^{-j2\pi kn/L},
\label{eq:dft}
\end{align}
after which the DC component is removed. 
Magnitude values are computed from the complex DFT output, 
then undergo normalization and logarithmic scaling,
\begin{align}
B[k] = 10\log_{10}(|X[k]|/L),
\label{eq:db_conversion}
\end{align}
%
To reduce complexity and sensitivity to narrow-band switching harmonics, adjacent frequency bins are aggregated,
\begin{align}
B_s[m] = \sum_{k=mS}^{(m+1)S-1} B[k],\qquad S=2.
\label{eq:band_aggregation}
\end{align}
The resulting $L/{2S}$-dimensional vector is used as the input to LD-Spec.

\subsubsection{LD-Spec Architecture}
\begin{figure}[!htbp]
\centering
\includegraphics[width=0.49\textwidth]{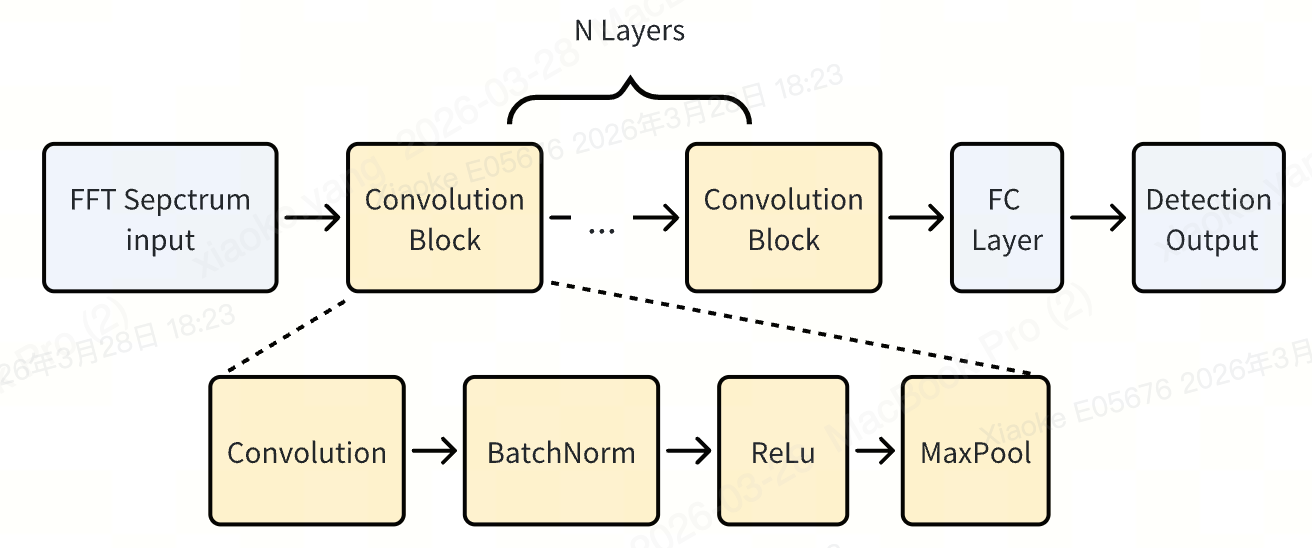}
\caption{Architecture of the spectral CNN (LD-Spec) 
for lightweight on-device arc-fault detection.}
\label{fig:ld_specnet_architecture}
\end{figure}

To extract discriminative spectral features under the computational and memory constraints of inverter platforms, we design \textit{LD-Spec}, a compact convolutional neural network employing kernels that operate exclusively along the frequency axis. This architecture reflects the physical nature of DC arc spectra, which are characterized by broadband stochastic patterns rather than time-localized transients. As illustrated in  Fig.~\ref{fig:ld_specnet_architecture}, LD-Spec processes the spectral input vector using a series of lightweight convolutional blocks. Each block consists of convolution, batch normalization, ReLU activation, dropout ($p=0.2$), and max-pooling. A fully connected layer with a classification head then outputs the likelihood of arc $P_{\mathrm{arc}}$.

\subsubsection{Dual-Threshold Decision Mechanism}
On top of the frame-level decision threshold of the classification head, the system employs a counter to track consecutive arc-positive frames, in order to suppress false alarms caused by transient disturbances. An arc alarm is triggered only when the counter value exceeds a predefined threshold $C$. The maximum threshold is bounded by detection latency specified by certification standards, and the optimal value is tuned in laboratory tests to achieve a target false alarm rate.

\subsection{LD-Align: Cross-Hardware Transfer}
\label{subsec:LD-Align}



LD-Align is designed to preserve the structural conditions required for near-perfect AFCI performance across heterogeneous hardware platforms. It transforms cross-converter deployment from a data-intensive retraining process into a transfer procedure that enables reliable baseline evaluation on new devices with minimal additional data.

\subsubsection{Feature Construction}
LD-Align transfer learning employs identical feature input for both source and target domains, specifically the Hanning-windowed FFT spectrum defined in Eq.~\eqref{eq:frame_segmentation}--\eqref{eq:dft}. Given that cross-converter shifts appear primarily as variations in harmonic magnitude and broadband noise levels, the FFT-based feature sufficiently captures the discrepancies required for alignment. Consequently, this design preserves a lightweight inference pipeline compatible with deployment constraints.

\subsubsection{Transferable Representation Learning and Alignment Loss Design}
The extracted spectral features are processed by the same compact convolutional classifier in LD-Spec, consisting of a backbone $G(\cdot)$ and a linear layer $C(\cdot)$.
LD-Align performs mixed-domain fine-tuning using target mini-batches together with source replay, optimizing the composite objective
\begin{align}
\mathcal{L}_{\mathrm{total}}
=
\alpha\,\mathcal{L}_{\mathrm{tgt}}
+
\beta\,\mathcal{L}_{\mathrm{src}}
+
\lambda\mathcal{L}_{\mathrm{L2}}.
\label{eq:ldalign_total_loss}
\end{align}
where $\alpha$, $\beta$, and $\gamma$ are hyperparameters. The source loss weight $\beta$ is fixed to 1 as baseline, $\alpha$ and $\gamma$ are are optimized via grid search. $\alpha$ balances target    supervision and $\gamma$ prevents overfitting. Each of the loss terms is discussed in detail below.

\paragraph{Target-domain supervision}

The target-domain loss
\begin{align}
\mathcal{L}_{\mathrm{tgt}}
=
\frac{1}{N_t}
\sum_j
\ell\!\left(
C(G(x_j^{(t)})),
y_j^{(t)}
\right)
\label{eq:ldalign_tgt_loss}
\end{align}
introduces hardware-specific supervision to correct mismatch-induced errors while maintaining the original decision structure. Target samples are collected with the same procedure as source samples. In the absence of target samples, the total loss reduces to the regularized source loss, and the resulting model degenerates to one trained solely on the source domain. Further $N_t$ is much smaller than samples in the source domain $N_s$, thus the data acquisition load is much smaller for the target domain.

\paragraph{Source replay for stability}

To mitigate catastrophic drift, LD-Align mixes each target update with replayed source mini-batches and minimizes the source classification loss 
\begin{align}
\mathcal{L}_{\mathrm{src}}
=
\frac{1}{N_s}
\sum_i
\ell\!\left(
C(G(x_i^{(s)})),
y_i^{(s)}
\right),
\label{eq:ldalign_src_loss}
\end{align}
which constrains the adapted model to preserve the high-performance source-domain decision boundary.

\paragraph{L2 regularization}
To further preserve feature stability , LD-Align applies L2 regularization to the trainable parameters, 
\begin{align}
\mathcal{L}_{\mathrm{L2\text{-}SP}}
=
\sum_{w\in\mathcal{W}_{\mathrm{train}}}
\bigl\|
w - w^{(0)}
\bigr\|_2^2,
\label{eq:ldalign_l2sp}
\end{align}
where $w^{(0)}$ denotes the pre-trained LD-Spec network weights and $\mathcal{W}_{\mathrm{train}}$ represents parameters updated during adaptation.

\paragraph{Layer-wise learning-rate scheduling}

The backbone is fine-tuned with a smaller learning rate than the classification
head.
Learning rates are reduced when validation macro-F1 saturates, and early
stopping prevents over-adaptation.





With LD-Align enabled, mixed-domain fine-tuning restores latent spectral features and stabilizes frame-level predictions, enabling high performance detection even after cross-hardware transfer.

\subsection{LD-Adapt: Self-Adaptive Model Update}
\label{subsec:LD-Adapt}

While LD-Spec provides accurate on-device detection and LD-Align enables cross-converter transfer, LD-Adapt addresses long-term field changes.
%
It introduces a cloud-device coordinated self-adaptation mechanism that detects novel operating conditions and incorporates them via controlled updates, thereby maintaining long-term AFCI reliability under real-world conditions.

\subsubsection{Device-to-Cloud Novelty Detection}

On the device side, whenever the embedded detector identifies a potential arc alarm, the corresponding frame of current measurement, along with contextual operating conditions, is transmitted to the cloud via the blackbox. Crucially, a verification mechanism is implemented to distinguish between true arc faults and false positives. In initial deployment, fault classifications are validated by power conversion system experts who analyze operating conditions (voltage, current, load profile) against domain-specific rules. End-user confirmation is  also collected to ascertain ground truth. Expert decisions are logged to build a labeled case repository. To address scalability constraints, expert validation is transitioned to a rule-based expert system. The system verifies physical constraints (signal magnitude, frequency content, duration), operational context (load type, system configuration), and produces confidence scoring based on rule match strength. An experience replay buffer stores edge cases where rule confidence is low, enabling periodic rule refinement and expert review of ambiguous cases.
Based on the verification result, confirmed arc events are logged for archival purposes, indicating the model remains valid for this scenario. Conversely, confirmed false alarms are flagged for model adaptation to prevent recurrence.


\subsubsection{Two-Stage Adaptive Evolution}
When a batch of novel data samples is collected, LD-Adapt initiates a two-stage evolution process.
\paragraph{Stage~1: Hyperparameter and Model Parameter Evolution.}
In the first stage, the LD-Spec/LD-Align architecture remains unchanged. Grid search on learning-rate schedules, loss weights, and batch construction strategies is carried out to adapt the model to the new conditions. This lightweight stage enables rapid recovery without modifying model structure, memory footprint, or inference latency.

\paragraph{Stage~2: FLOPs-Bounded Micro-Architectural Evolution.}
If performance improvements from Stage~1 saturate, defined as the validation loss improvement falling below a predefined 0.5\% tolerance over 3 consecutive epochs, LD-Adapt activates a second stage that allows limited structural refinement. Allowed adjustments include kernel size and channel width tuning, minor fully connected layer modifications, and dropout rate optimization. 
All modifications are constrained to less than $5\%$ additional FLOPs, ensuring that real-time inference latency and MCU compatibility remain unchanged. Together, the two stages provide a safe mechanism for incorporating new operating conditions while preserving deployment constraints.

\subsubsection{Canary Deployment}
Candidate models undergo temporal validation followed by controlled canary  deployment. The updated model is first deployed to a small subset of devices, where the number of alarms and nuisance tripping rates are continuously monitored. Only when the new model consistently outperforms the existing baseline is it promoted to fleet-wide deployment. This canary mechanism ensures that unsafe updates are automatically contained and that AFCI reliability does not degrade during long-term operation. Overall, the cloud-device adaptive loop enables incremental integration of novel operating conditions while preventing error accumulation and maintaining stable AFCI protection in evolving PV-BESS environments.

\section{Experiments}
\label{sec:experiments}

To rigorously evaluate the robustness and real-world applicability of the proposed arc-fault detection framework, this section presents a comprehensive experimental study in various PV and BESS scenarios.
The evaluation is organized around two complementary objectives:
(1) assessing the robustness to nuisance trips under diverse normal operating conditions, and (2) quantifying detection performance under controlled series arc-fault scenarios that comply with industrial safety standards.
Collectively, these experiments validate the detection accuracy and operational reliability required for residential deployment.

\subsection{Test Scenario Construction and Dataset}

\begin{figure}[t!]
\centering
\includegraphics[width=0.85\linewidth]{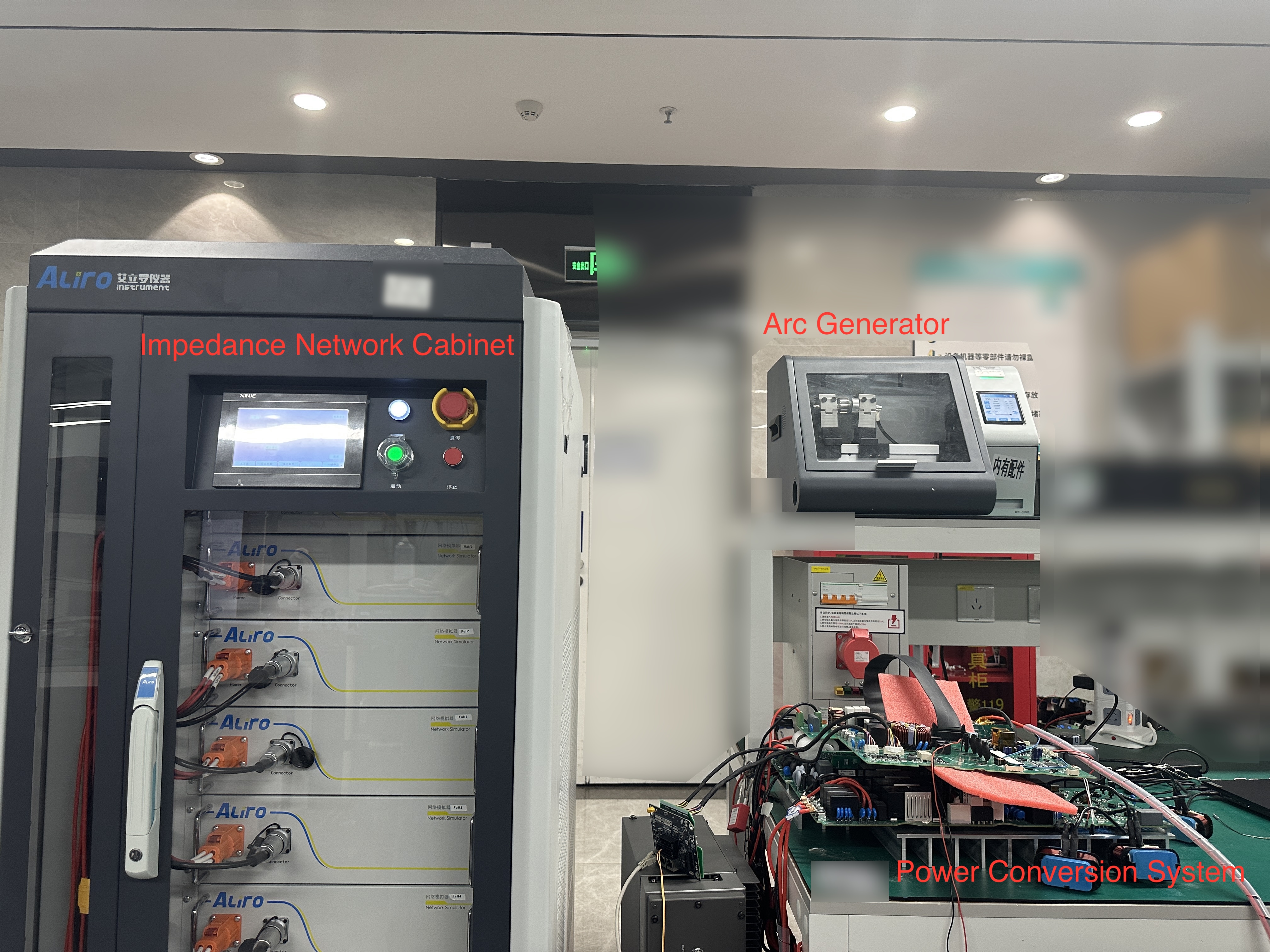}
\caption{Experimental data acquisition and testing environment consisting of an impedance network cabinet (left), an arc generator (upper right), and the power conversion system (PCS, lower right).}
\label{fig:data_acquisition_env}
\end{figure}

The test scenarios are designed to systematically capture the wide range 
of transient behaviors, electrical disturbances, and operating-mode 
transitions commonly observed in residential PV systems, or conditions 
known to trigger false trips in conventional AFCI devices. Complementing 
these nuisance-trip evaluations, controlled arc-fault data are collected 
under standardized fault scenarios. The laboratory setup, including the impedance network, power conversion system, and a controlled arc generator, is illustrated in Fig.~\ref{fig:data_acquisition_env}. Two power conversion systems are used in the experiments, with exactly the same testing configurations. Details of the hardware platforms are discussed in the transfer experiments. 

\begin{figure}[htbp!]
\centering
\begin{subfigure}{0.95\linewidth}
  \includegraphics[width=\linewidth]{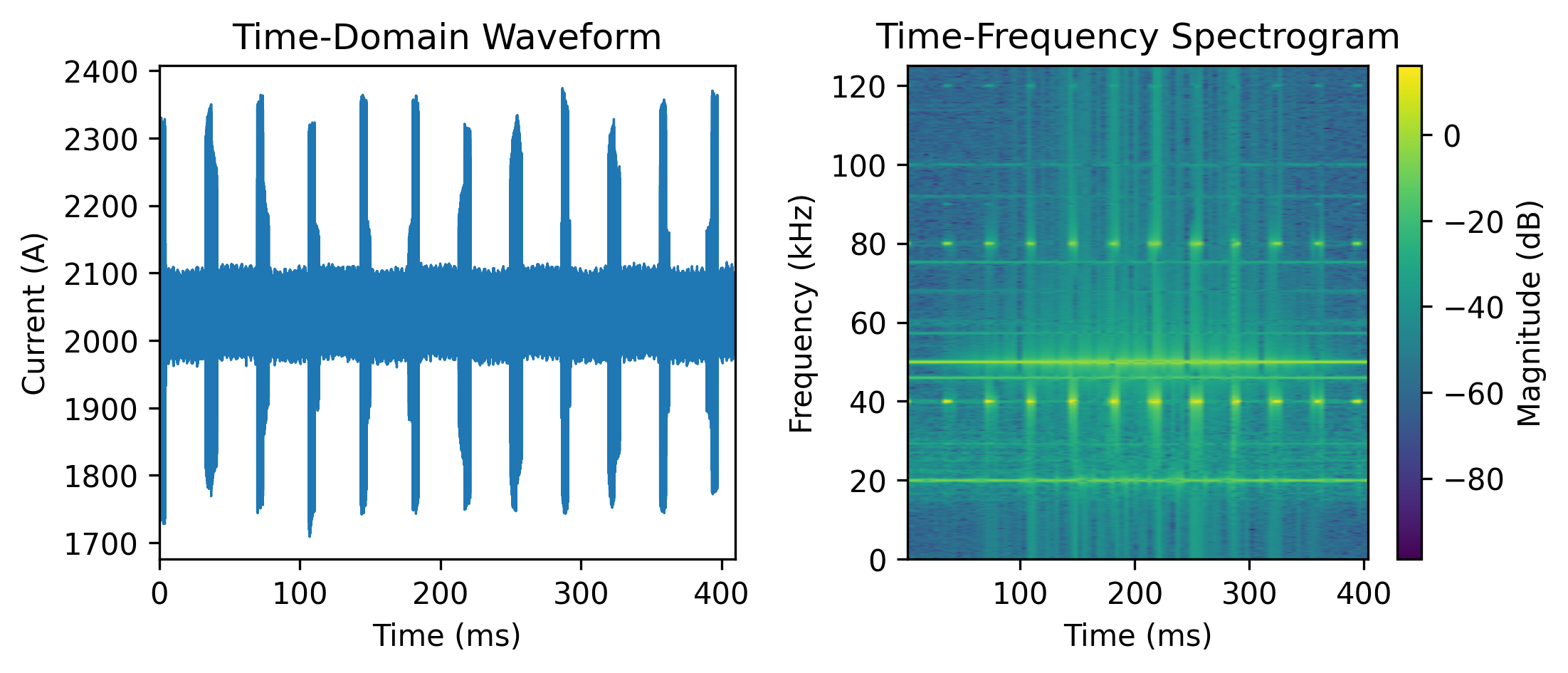}
  \caption{(a) PV current under no-load steady operation.}
\end{subfigure}
\vspace{0.35cm}
\begin{subfigure}{0.95\linewidth}
  \includegraphics[width=\linewidth]{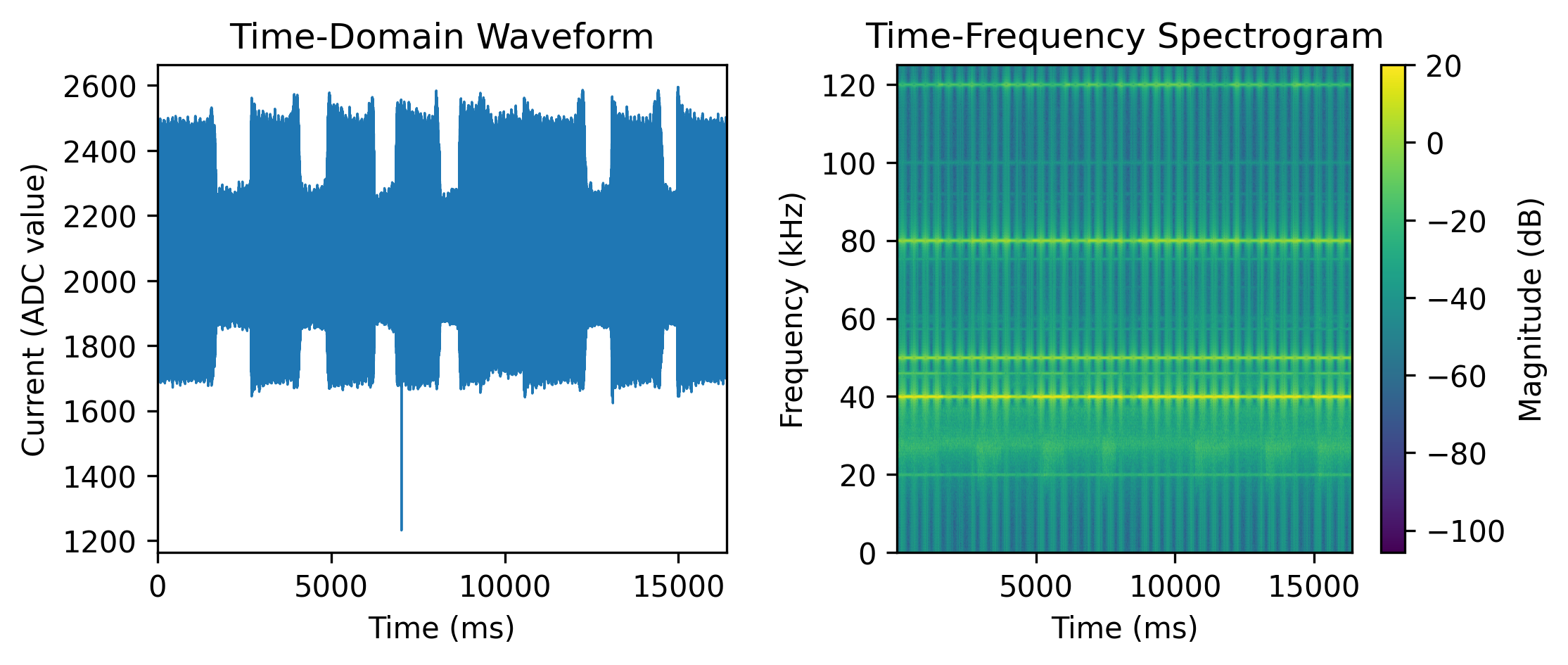}
  \caption{(c) PV current under parallel operation with power jump.}
\end{subfigure}
\caption{Spectrograms of representative operating conditions for nuisance trip evaluation.}
\label{fig:operating_conditions}
\end{figure}

To evaluate robustness against nuisance tripping, we constructed a comprehensive set of representative non-fault operating conditions. These scenarios cover transient disturbances, power-quality degradations, grid-interaction behaviors, and other conditions frequently encountered in real-world PV and BESS deployments. The test cases are grouped into the categories listed in Table~\ref{tab:operating_conditions}, with the number of detailed operating conditions specified for each. Arc-fault data are also collected across diverse operating conditions involving different PV currents and voltages. Figure~\ref{fig:operating_conditions} shows the spectrograms of representative operating conditions for nuisance trip evaluation. As can be seen, the spectrograms exhibits high frequency components at some time steps, which are likely to trigger nuisance trips.

\begin{table}[htbp]
\centering
\caption{Operating Conditions for Nuisance Trip Evaluation}
\label{tab:operating_conditions}
\begin{tabular}{p{5.5cm}p{2cm}}
\toprule
\textbf{Category} & \textbf{Operating Conditions} \\
\midrule
System Start-up Processes & 7 \\
Parallel Operation of PV Strings & 3 \\
PV Direct-Connection Modes & 6 \\
DC Circuit Breaker Operation & 4 \\
Variable PV Input Modes & 3 \\
Limited Grid-Feeding \& Repeated Start--Stop & 2 \\
Grid Connection and Disconnection & 4 \\
AC-Side Load Switching & 4 \\
Harmonic Grid Conditions & 2 \\
\bottomrule
\end{tabular}
\end{table}

\subsection{LD-Spec}
\label{subsec:LD-Spec-exp}

\subsubsection{Experimental Validation across All Operating Conditions}
The LD-Spec model was trained using the collected dataset. Training employed multiple folds of random train-test splits to ensure robust evaluation. The resulting performance metrics on the test sets are: Accuracy = 0.9999, Precision = 0.9996, Recall = 0.9996, and F1-score = 0.9996. Both ROC-AUC and PR-AUC exceed 0.999, indicating near-perfect separability between arc and non-arc conditions. 

The trained model was then deployed to the embedded device. Across all nuisance-trip-prone operating conditions, the proposed AFCI achieved a 0\% false-trip rate with 100\% arc fault detection, demonstrating robust immunity to diverse non-fault disturbances and validating its suitability for practical PV-BESS deployment. 

\begin{figure}[htbp!]
\centering
\subcaptionbox{(a) Arc fault detection: normal operation transitioning to arc event.}{%
  \includegraphics[width=0.88\linewidth]{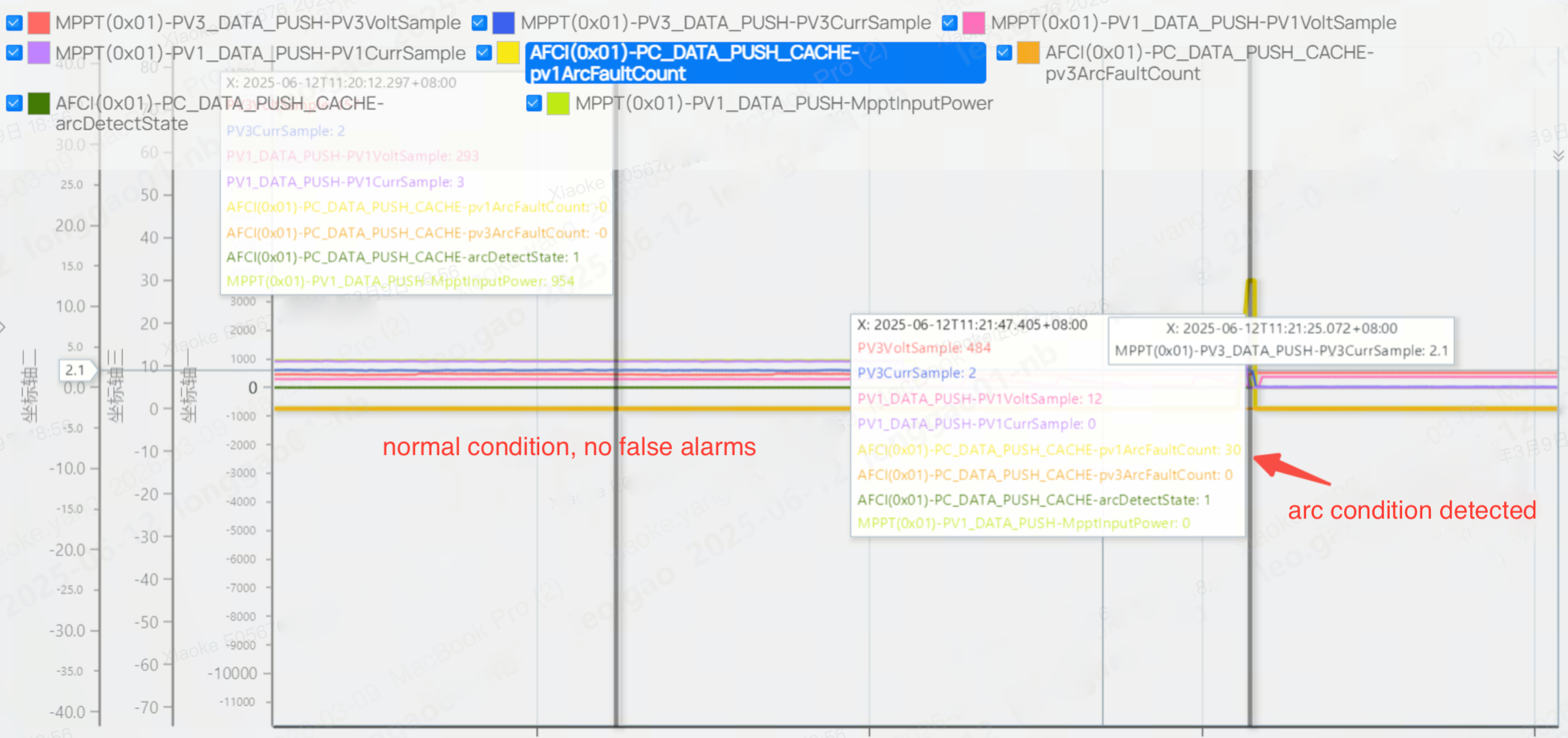}
  \label{subfig:ecomatrix_normal_and_arc}
}\\[1em]
\subcaptionbox{(b) Normal operation with detection spikes (no alarm triggered).}{%
  \includegraphics[width=0.88\linewidth]{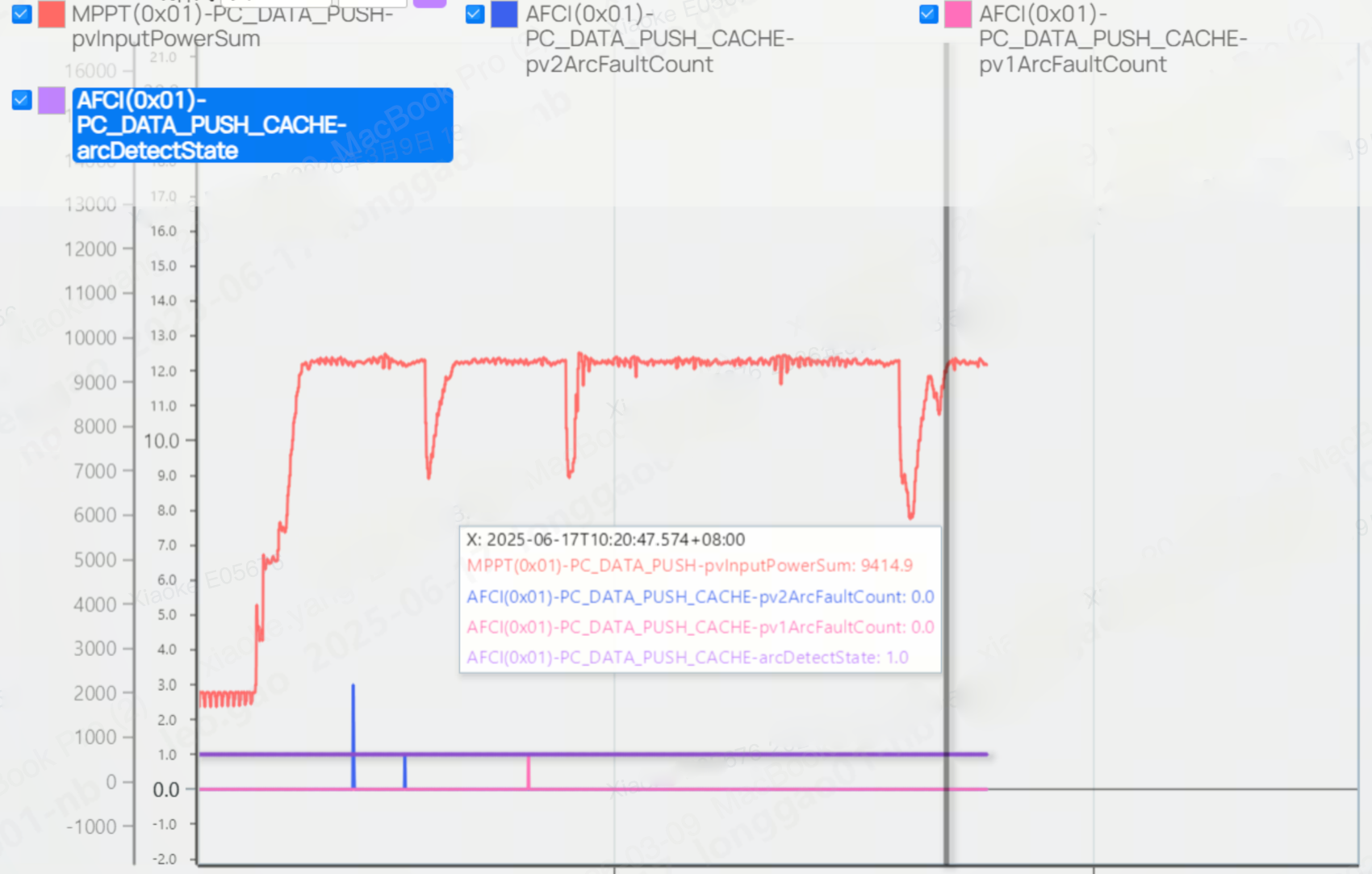}
  \label{subfig:ecomatrix_nromal}
}
\caption{Screenshots of some test cases of the proposed AFCI method.}
\label{fig:ecomatrix}
\end{figure}

Figure~\ref{fig:ecomatrix} shows two test scenarios. In Fig.~\ref{fig:ecomatrix}(a), the system initially operates under normal conditions with no AFCI alarms. When an arc fault is introduced, the counter rapidly reaches the threshold (indicated by 
the yellow line), immediately triggering an alarm. In contrast, Fig.~\ref{fig:ecomatrix}(b) depicts a normal operating condition where the AFCI counters (blue line and pink line for two channels of PV input) exhibit several spikes, indicating positive detection of some frames. However, since the counter never reaches the threshold, no alarm is triggered. This demonstrates the effectiveness of the dual-threshold mechanism in distinguishing true arc faults from transient disturbances.

\begin{figure}[htbp!]
\centering
\subcaptionbox{(a) Arc fault at PV string START}{%
  \includegraphics[width=0.88\linewidth]{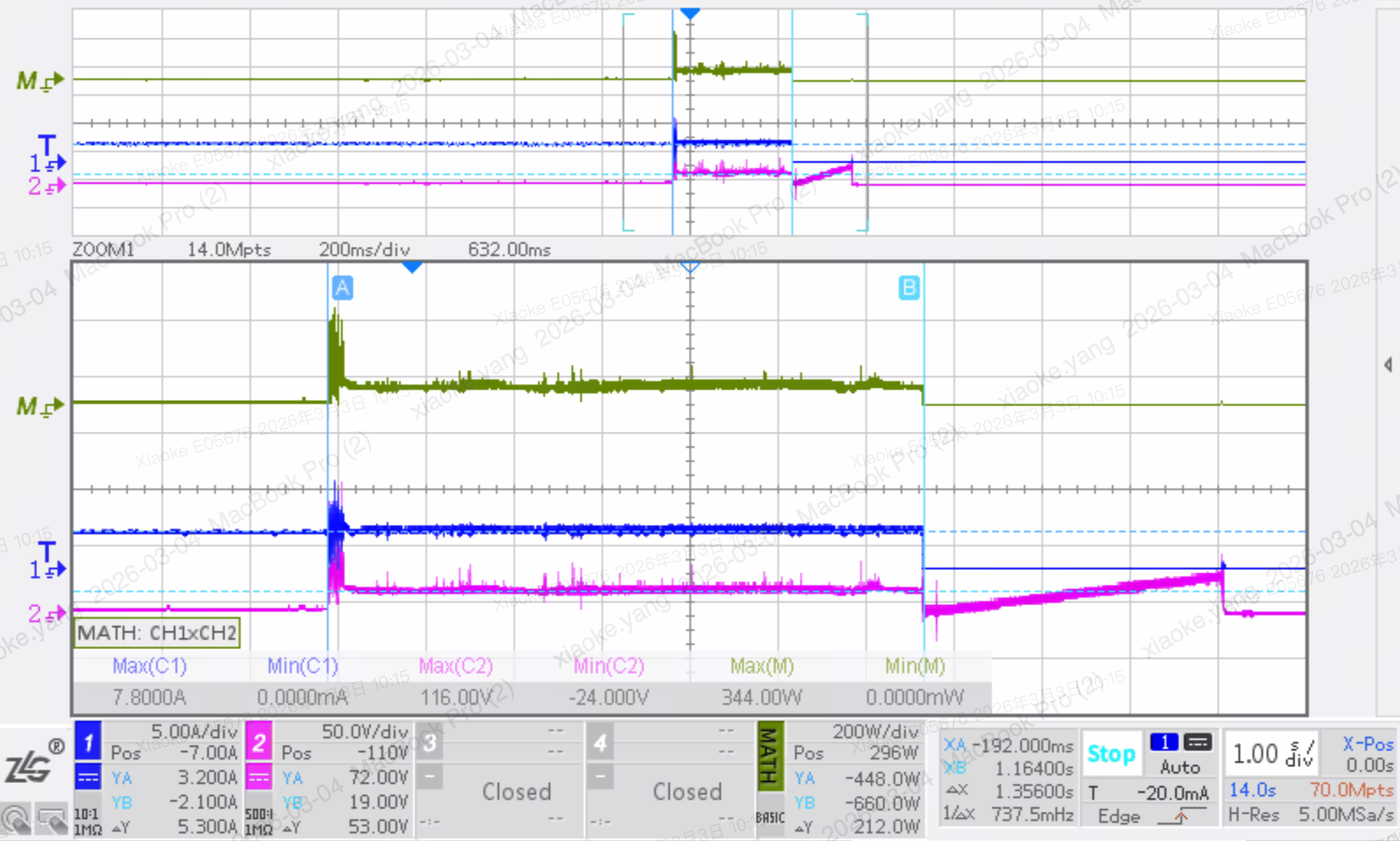}
  \label{subfig:ul_start}
}\\[1em]
\subcaptionbox{(b) Arc fault at PV string MIDDLE}{%
  \includegraphics[width=0.88\linewidth]{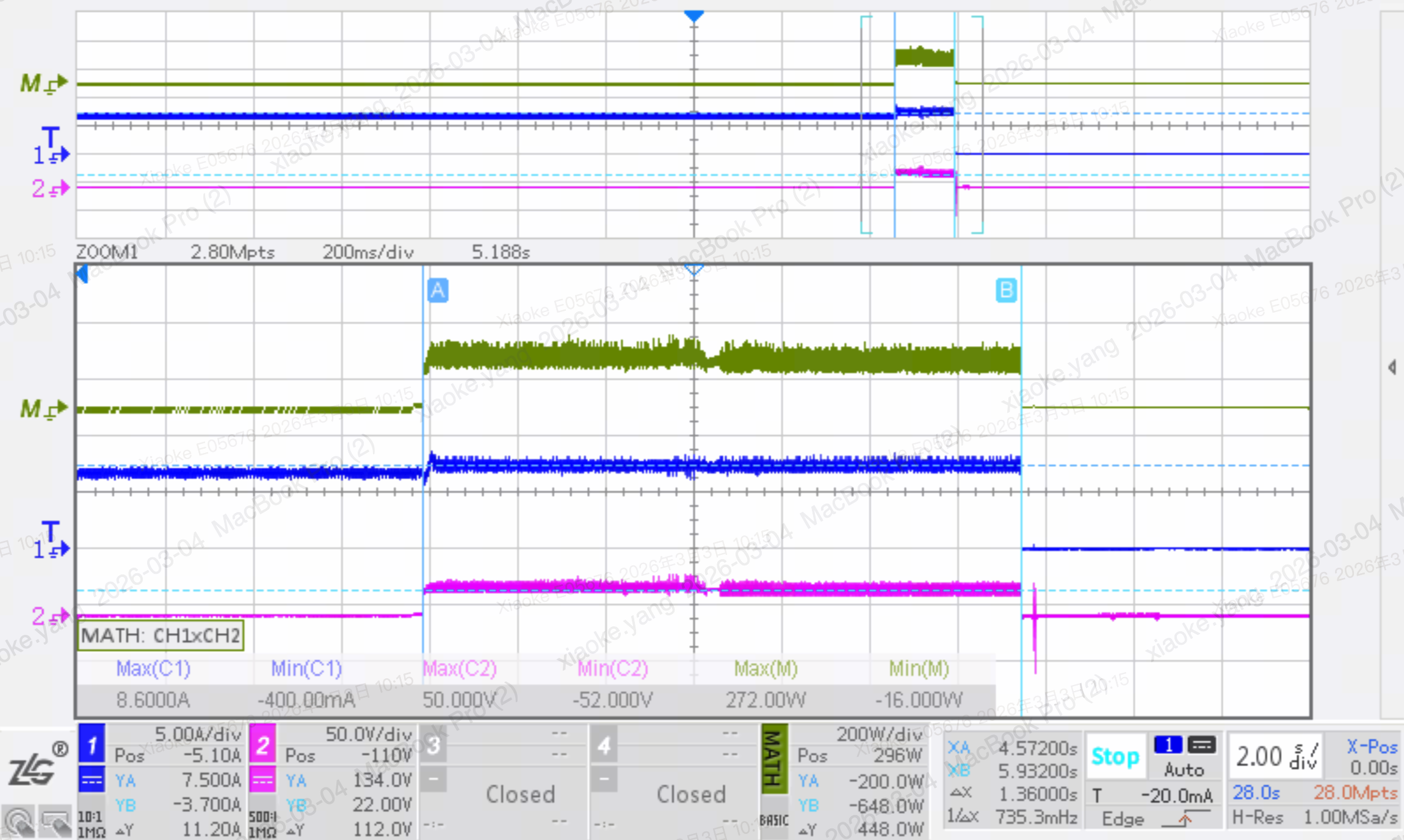}
  \label{subfig:ul_middle}
}
\caption{Oscilloscope waveforms during series arc fault tests on a PV string: 
current (Ch1), voltage (Ch2), and calculated power (MATH). The fault clearance time $\Delta$ x is measured from arc initiation to AFCI trip. For the START location (a), the time is 1.356~s; for the MIDDLE location (b), the time is 1.36~s.}
\label{fig:ul_tests}
\end{figure}

Furthermore, an impedance network cabinet was employed to simulate the test configurations specified in UL~1699B standards. Representative oscilloscope waveforms captured during UL testing are shown in Fig.~\ref{fig:ul_tests}. The figure demonstrates that rapid PV shutdown is triggered upon arc detection before the time and energy thresholds required by safety standards are reached. 

\subsubsection{Scaling Analysis and Data-Saturation Behavior}

\begin{figure}[htbp!]
\centering
\includegraphics[width=0.99\linewidth]{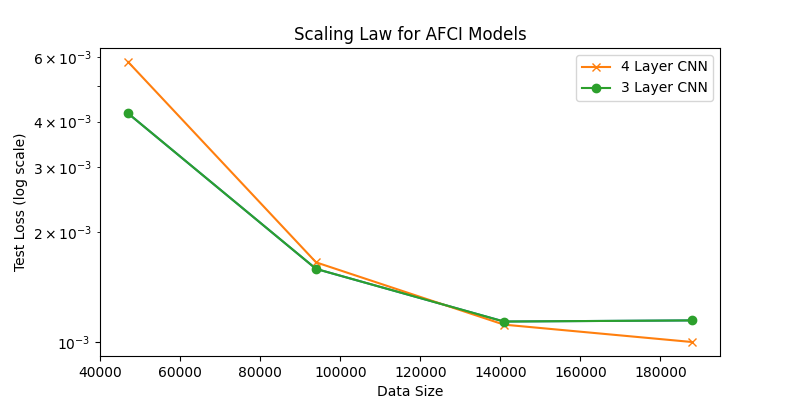}
\caption{Scaling behavior of LD-Spec under increasing dataset sizes.}
\label{fig:ld_specnet_scaling_law}
\end{figure}

To quantify the amount of data required for reliable embedded deployment, we
conducted a scaling analysis by training LD-Spec on progressively larger
subsets of the full dataset. The model exhibits a clear power-law trend in test
loss:
\begin{equation}
L(N)=aN^{-\alpha}+L_\infty, \qquad \alpha \approx 0.37,
\end{equation}
where $L(N)$ is the test loss at data size $N$ and $L_{\infty}$ the asymptotic error. Loss decreases monotonically up to 140 k samples, beyond which the slope flattens, signalling the onset of capacity saturation. To further expose the parameter–data trade-off, we compared a 3-layer (4.6 k) and a 4-layer (6.8 k) CNN on the same 230 k corpus, as shown in Figure~\ref{fig:ld_specnet_scaling_law}. 

Extrapolating the 4-layer curve shows that an additional 220 k samples would shave only 0.0001 off the current 0.001 0 loss, and the marginal return drops below 5\%. For the 4-layer CNN, 230 k samples already satisfy this criterion; 
Extrapolation suggests that expanding the sample pool by another order of magnitude would reduce test loss only marginally. This provides a practical guideline for determining the size of real-world data required to get a reasonable LD-Spec model. 


\subsection{LD-Align: Cross-Hardware Transfer Experiments}
\label{subsec:ld_align_experiments}
In practical AFCI deployments, arc-fault detection models are typically trained on one inverter platform but must generalize to different platforms with distinct electrical characteristics. To evaluate the robustness of LD-Align under such heterogeneous hardware conditions, we conduct cross-hardware transfer experiments over two hardware platforms. 

\begin{figure*}[t]
\centering
\begin{subfigure}[t]{0.32\textwidth}
    \centering
    \includegraphics[width=\linewidth]{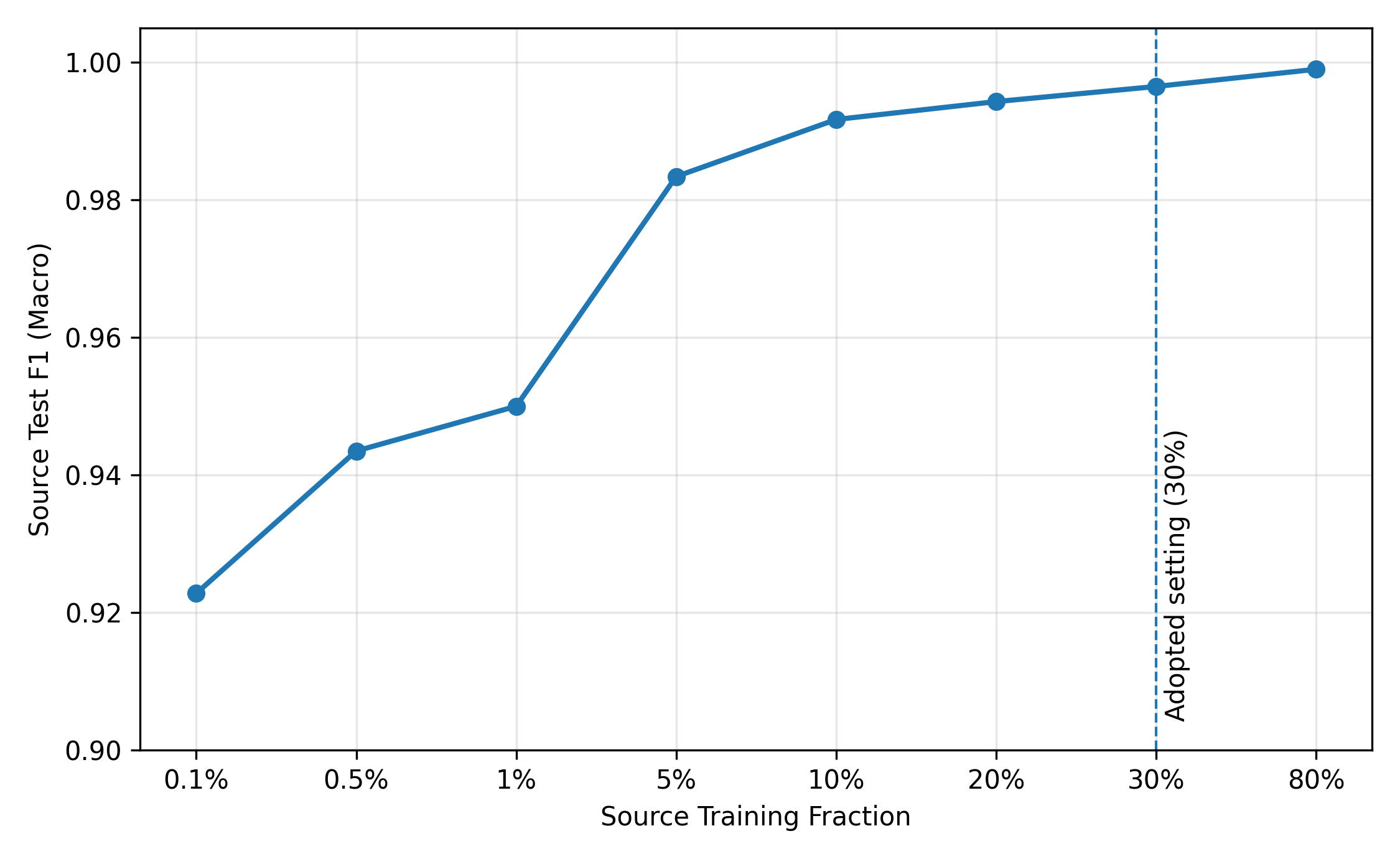}
    \caption{(a) Source macro-F1 vs. fraction of source training data. Source performance saturates beyond 30\% supervision.}
    \label{fig:sub_source_fraction}
\end{subfigure}
\hfill
\begin{subfigure}[t]{0.32\textwidth}
    \centering
    \includegraphics[width=\linewidth]{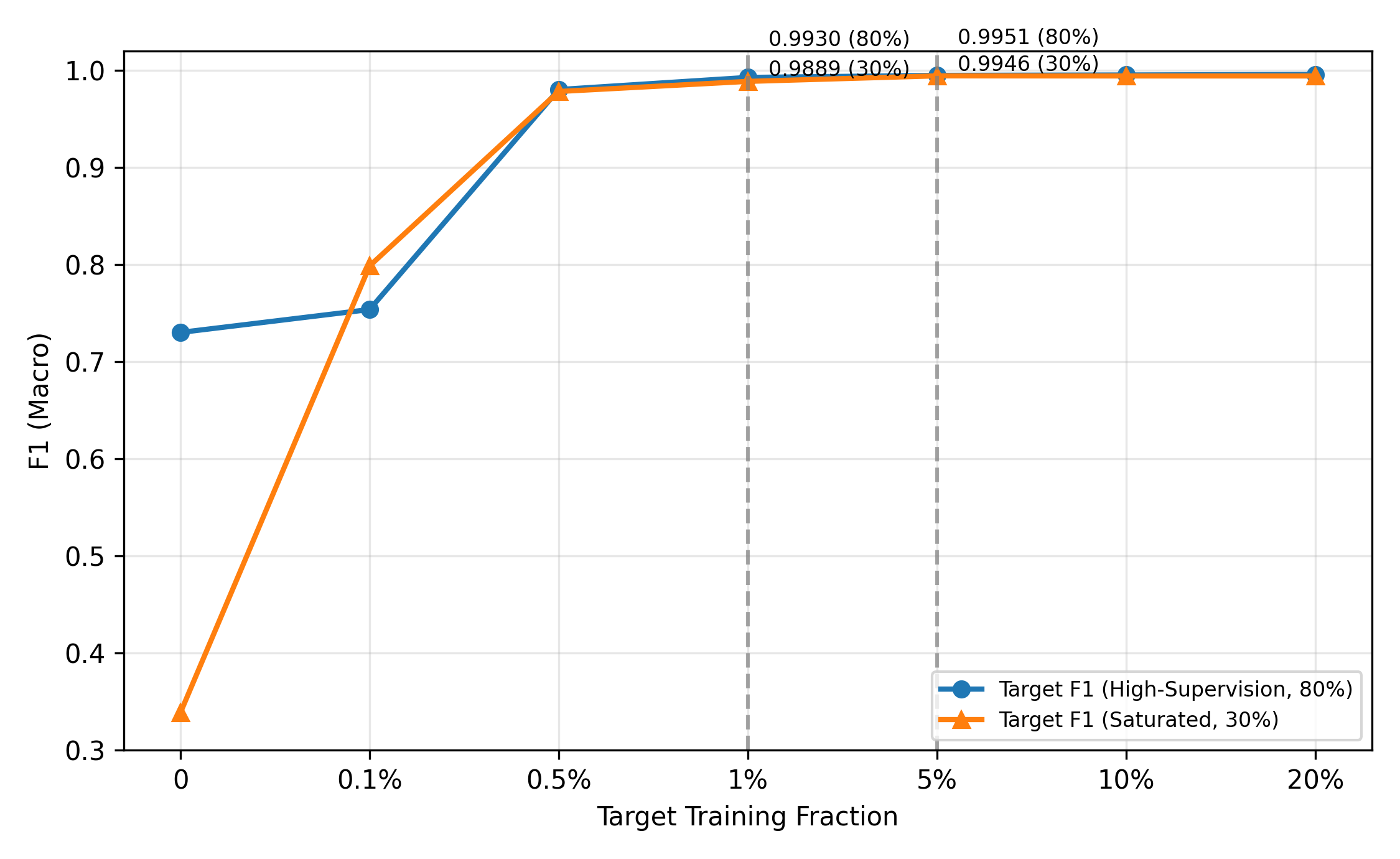}
    \caption{(b) Target macro-F1 vs. fraction of target training data. Two curves correspond to different fractions of source data used. Increasing target supervision improves macro-F1.}
    \label{fig:sub_target_sweep_overall}
\end{subfigure}
\hfill
\begin{subfigure}[t]{0.32\textwidth}
    \centering
    \includegraphics[width=\linewidth]{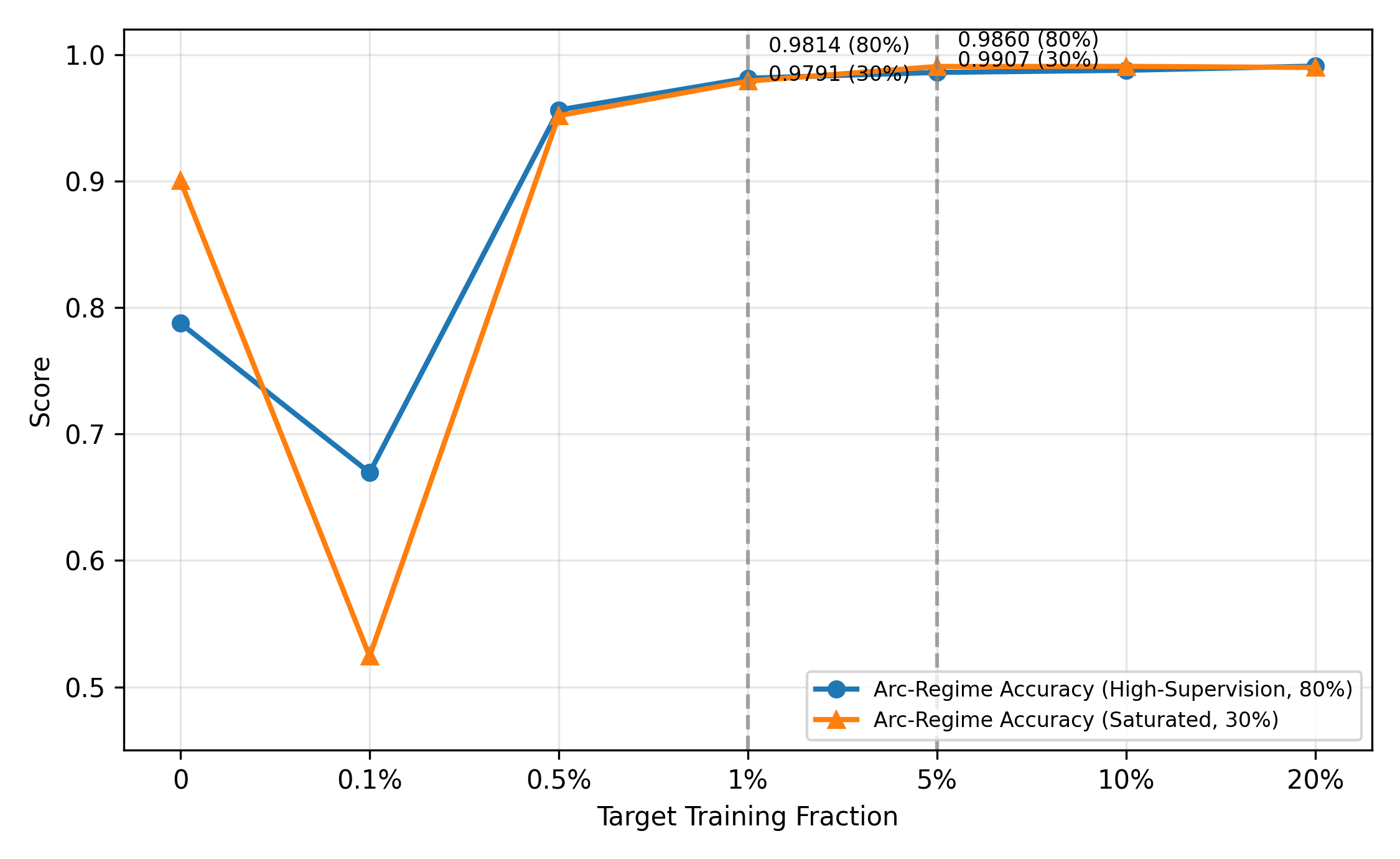}
    \caption{(c) Arc detection accuracy vs. fraction of target training data. Two curves correspond to different fractions of source data used and follow the same pattern as (b).}
    \label{fig:sub_target_sweep_arc}
\end{subfigure}
\caption{Cross-hardware transfer behavior of LD-Align.}
\label{fig:ld_align_source_target}
\end{figure*}

Although both platforms share identical PV current sampling configurations, including sampling rate, resolution, and sensor placement, they differ in key operational parameters. Most notably, the switching frequencies are different: the source platform operates at 40 kHz, while the target platform uses 25 kHz. These differences manifest as distinct spectral signatures, as illustrated in Fig.~\ref{fig:series_frequency_diff}. The transfer experiment results are summarized in Figure~\ref{fig:ld_align_source_target}.


\subsubsection{Experiment I: Source Model}
During source model training, we collected a dataset of 230,000 samples from the source inverter platform. To establish a reliable baseline for cross-hardware transfer, we first determine the fraction of training data, or supervision level required to obtain a source model suitable for transfer. As shown in Fig.~\ref{fig:ld_align_source_target}(a), source macro-F1 increases rapidly with 5\%--10\% supervision data and exhibits clear diminishing returns thereafter. Performance saturates around 30\% supervision, increasing the training data from 30\% to 80\% improves macro-F1 by less than 0.3\%. Based on this saturation behavior, we select two representative models for transfer evaluation: (i) a high-supervision model (with 80\% data) and (ii) a saturated model (with 30\% data).
\subsubsection{Experiment II: Target Model Adaptation}
To enable rigorous evaluation of target-domain adaptation, we additionally collected a dedicated dataset comprising 218,000 labeled samples from the target inverter platform. This target dataset serves two purposes: (1) establishing an upper-bound performance reference via full-supervision training, and (2) enabling controlled ablation studies of adaptation efficiency under limited labeling budgets. We evaluate target adaptation by varying the fraction of labeled target data while keeping the source model fixed. Fig.~\ref{fig:ld_align_source_target}(b) shows that target macro-F1 increases monotonically with target supervision and exhibits saturation around 0.5\%--1\%. Beyond this region, additional supervision yields only marginal gains. Fig.~\ref{fig:ld_align_source_target}(c) further explores the arc-specific detection accuracy. The saturated 30\% model achieves nearly identical arc detection accuracy as the high supervision one once a small amount of target supervision is introduced. 

This results indicate that a fully supervised source model is not necessary for effective cross-hardware transfer. A moderately supervised source model already provides sufficiently transferable performances. Reduced source supervision also means lower effort for data acquisition.

\subsubsection{Empirical Transfer Law and Deployment Implication}

The experiments reveal a consistent empirical law:

\begin{center}
Target supervision $\uparrow$
$\Rightarrow$
Target performance $\uparrow$ (saturating),
Source performance $\downarrow$ (gradual).
\end{center}

This behavior holds for both macro-F1 and arc dectection accuracy. Combined with the strong saturation observed in source training, these results indicate that a moderately supervised source model (approximately 30\%) together with a small representative target subset (0.5\%--1\%) is sufficient to recover most target performance. This provides a practical and scalable deployment strategy for cross-hardware AFCI adaptation.

\subsection{LD-Adapt}
The LD-Adapt experiments aim to address the discrepancies between real-world and laboratory conditions. Two specific cases are examined: (1) differences between actual PV panels and simulated PV sources used in lab testing, as illustrated in Fig.~\ref{fig:pv_source_vs_panel_spectro}, and (2) environmental variations encountered during field deployment.

\paragraph{Performance Recovery on PV Panels} 
To validate the generalization capability, we evaluated the model on in-house rooftop PV panels under arc-fault conditions. Due to spectral discrepancies between the laboratory setup and field conditions, the baseline model achieved only 21\% precision. We then applied the \textit{LD-Adapt} Stage~1 adaptation mechanism, which performs incremental parameter fine-tuning using the newly collected PV panels data. Notably, this process utilized standard hyperparameter settings without requiring exhaustive search. After adaptation, the detection precision recovered to 95\%, confirming that parameter-level adjustment is sufficient to bridge the gap between simulated and physical PV sources.

\begin{figure}[htbp!]
\centering
\includegraphics[width=0.90\linewidth]{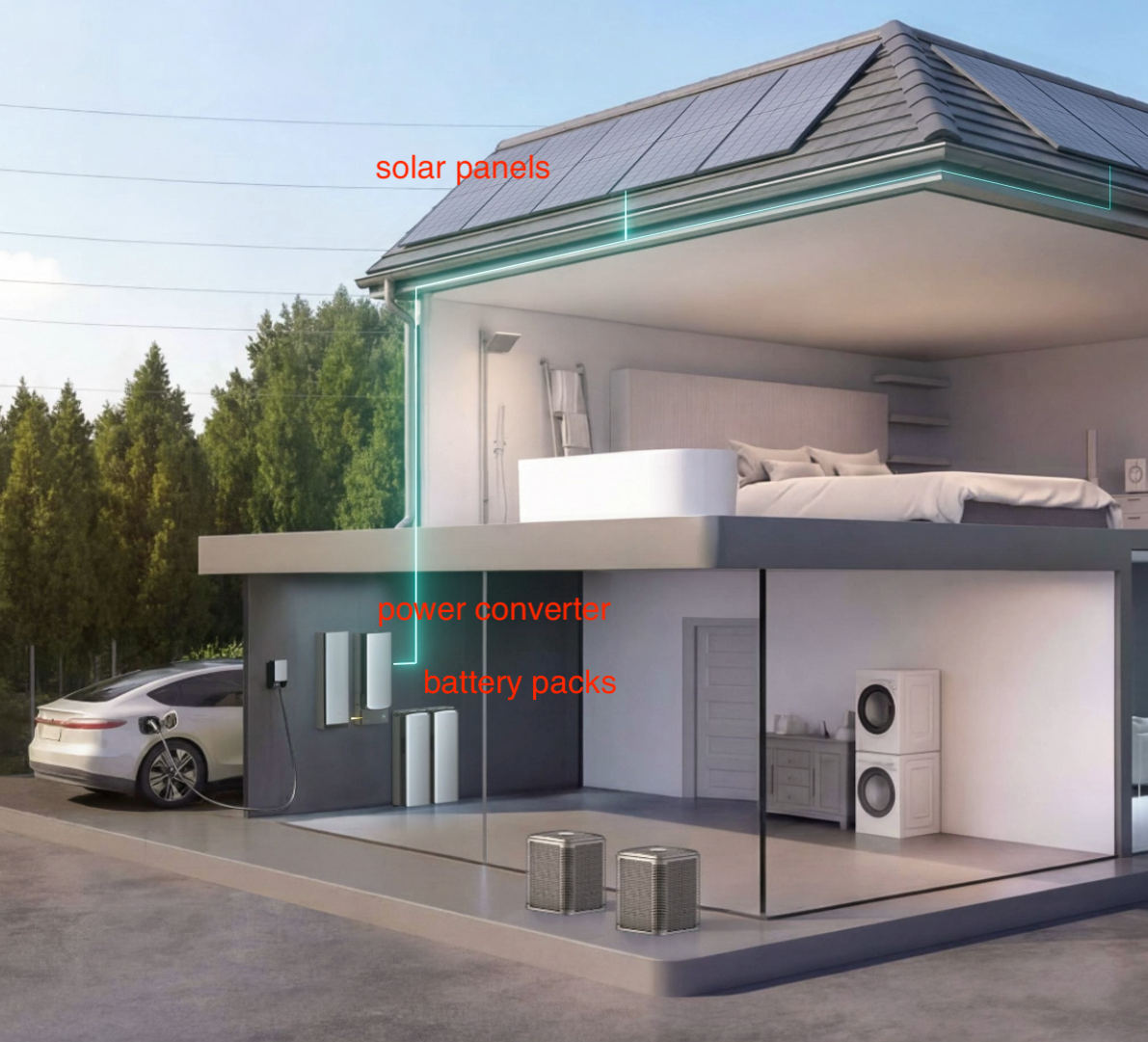}
\caption{Typical installation of the photovoltaic and battery energy storage system.}
\label{fig:field_installation}
\end{figure}

\paragraph{Performance Recovery on Field Conditions} 
A large fleet of devices has been deployed in the field, and typical installation is illustrated in Figure \ref{fig:field_installation}.  

\begin{figure}[htbp!]
\centering
\includegraphics[width=0.90\linewidth]{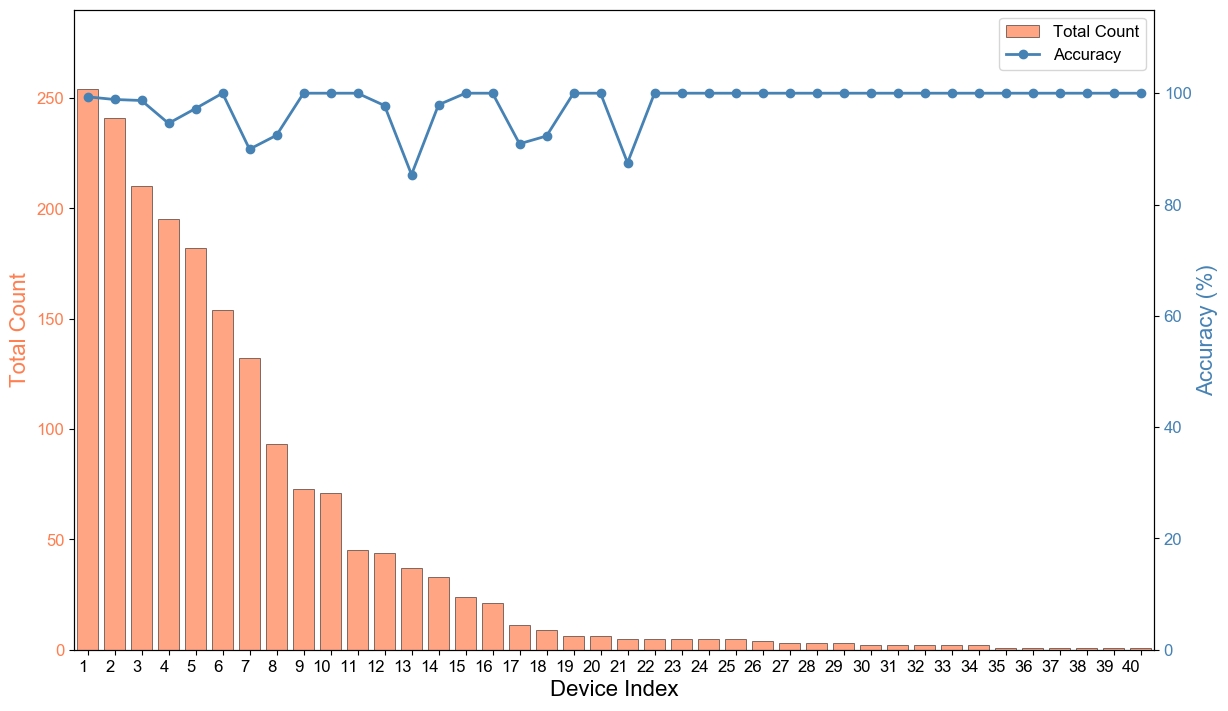}
\caption{AFCI alert statistics from field-deployed over a one-month period. Device Index denotes an individual device, with notable variation in false alarm rates across devices. The line on top shows detection precision when LD-Adapt is applied.}
\label{fig:real_device}
\end{figure}

\begin{figure}[htbp]
\centering
\begin{subfigure}{0.95\linewidth}
    \includegraphics[width=\linewidth]{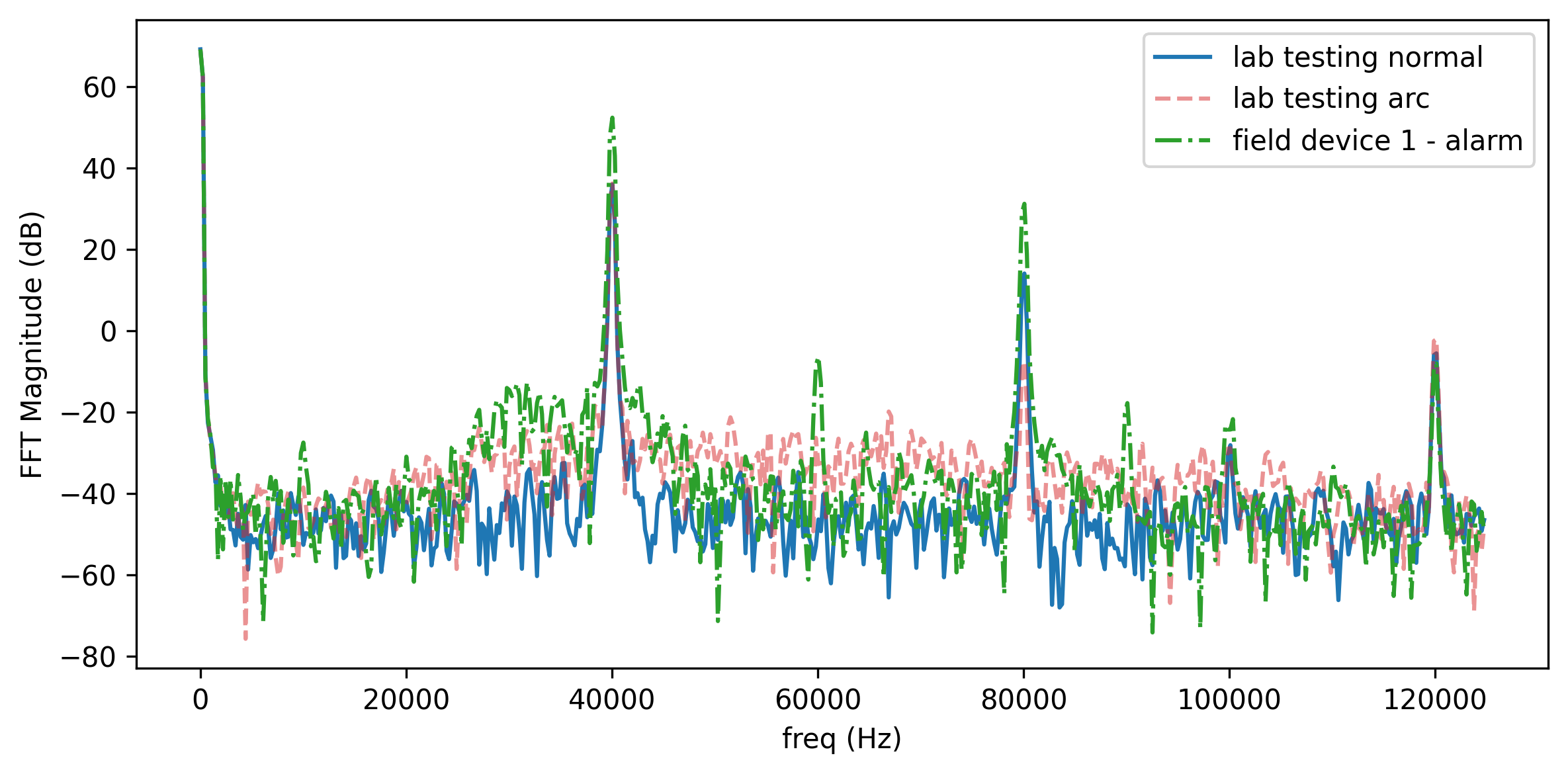}
    \caption{(a) Device 1}
\end{subfigure}
\vspace{0.35cm}
\begin{subfigure}{0.95\linewidth}
    \includegraphics[width=\linewidth]{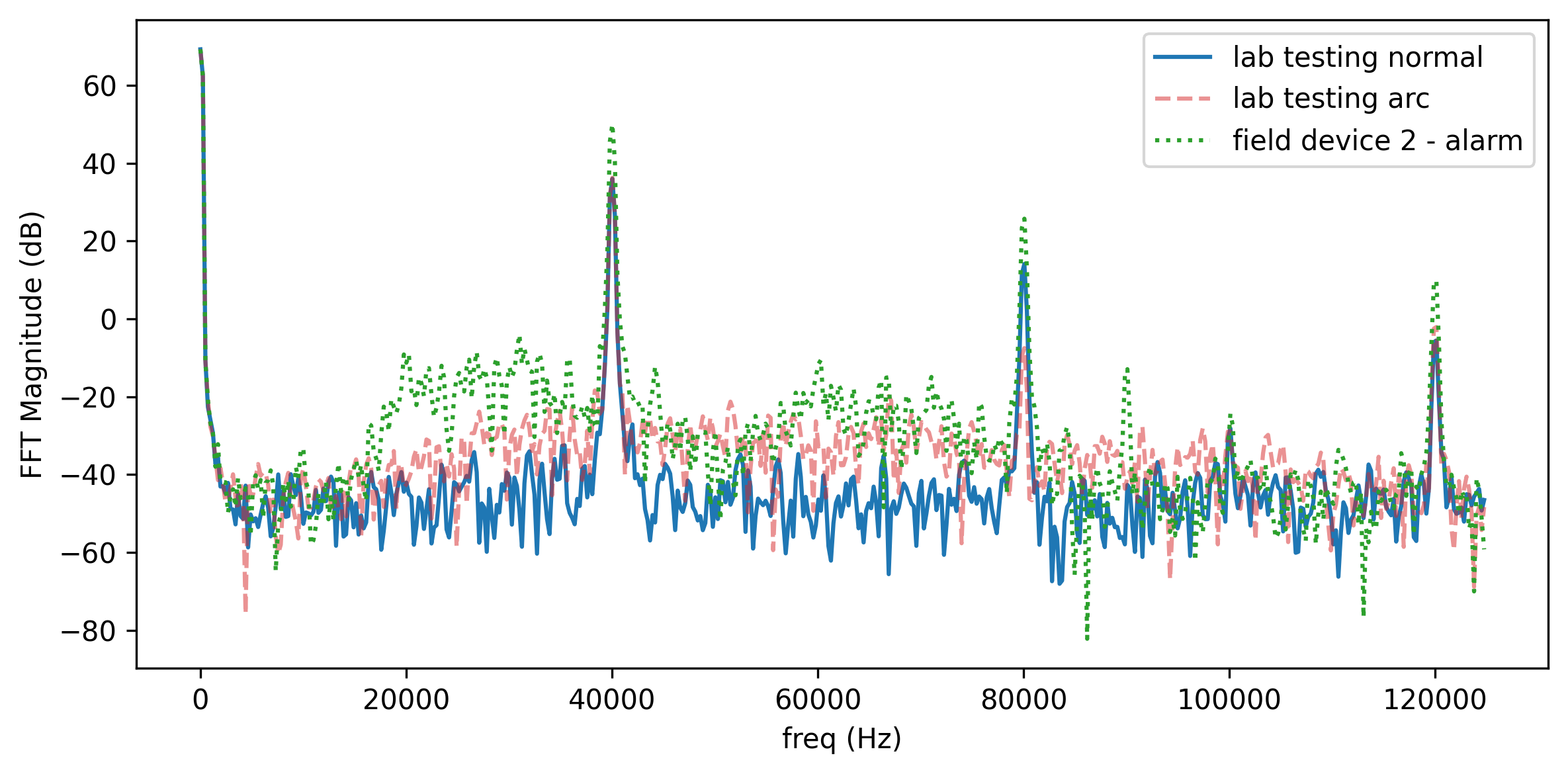}
    \caption{(b) Device 2}
\end{subfigure}
\caption{Spectra of false alarms compared with lab testing normal and arc conditions. Distinct spectral deviations highlight domain shift between field and lab testing environments.}
\label{fig:field_device_spectro}
\end{figure}

The algorithm performing reliably across the majority of units, yet AFCI alarms were triggered by a small subset of devices due to unique environmental conditions. Alarm frames were collected and filtered through the verification mechanism to identify confirmed false positives. Only these verified false alarms entered the cloud-device adaptation loop. Fig.~\ref{fig:real_device} shows the distribution of verified false alarms data collected from these devices and the recovered detection precision for each device after Stage~1 adaptation. Fig.~\ref{fig:field_device_spectro} compares alarm spectra from Devices~1 and 2, revealing distinct deviations from laboratory-tested nominal conditions. These spectral discrepancies are effectively handled by LD-Adapt method. In all observed cases, performance was restored without activating Stage~2 structural evolution, indicating that the domain shifts encountered were within the correction capacity of parameter fine-tuning. 

Across all adaptation experiments, the \textit{LD-Adapt} framework successfully restored performance using only Stage~1 parameter updates. The Stage~2 micro-architectural evolution (FLOPs-bounded structural 
refinement) was not triggered, as the performance saturation point was not reached during Stage~1. This validates the hierarchical design: lightweight parameter adaptation handles common drifts, while reserving 
structural changes for more severe shifts. Overall, the experiments confirm that LD-Adapt functions as a practical self-maintenance mechanism, enabling sustained, effectively 100\% reliable AFCI operation under long-term real-world evolution with minimal computational overhead.

\section{Conclusion}
\label{sec:conclusion}
This paper proposes a lightweight, transferable, and self-adaptive learning framework for DC arc-fault detection in residential PV-BESS systems. Motivated by practical challenges including inverter-induced spectral interference, cross-hardware heterogeneity, long-term operating-condition drift, and stringent microcontroller constraints, the proposed LD-framework is designed to sustain high AFCI performance under real-world deployment conditions.
The proposed framework integrates three complementary components. At the device level, \textit{LD-Spec} learns compact spectral representations that enable accurate and efficient on-device arc-fault detection within individual PV-BESS systems. Across heterogeneous converter platforms, \textit{LD-Align} preserves arc-discriminative spectral features in the presence of converter-specific discrepancies, maintaining feature consistency required for robust detection. Beyond static deployment, \textit{LD-Adapt} addresses long-term drift by incorporating novel operating conditions through a cloud-device coordinated self-evolution process. 
Overall, the proposed LD-framework advances AFCI technology from a static, laboratory-trained classifier toward a maintainable, transferable, and deployment-oriented learning-driven protection mechanism, providing a practical foundation for next-generation PV safety systems. Extensive experiments on both laboratory testbeds and real-world deployments demonstrate that the proposed framework achieves reliable arc-fault detection with zero nuisance-trip rates, rapid response, and robust generalization across diverse operating conditions, all without exhaustive retraining or excessive data collection. Overall, the proposed LD-framework advances AFCI technology from a static, laboratory-trained classifier toward a maintainable, transferable, and deployment-oriented learning-driven protection mechanism, providing a practical foundation for PV-BESS safety systems.

\ifCLASSOPTIONcaptionsoff
  \newpage
\fi
\bibliographystyle{IEEEtran}
\bibliography{IEEEab,testbib}
\end{document}